\documentclass[times,twocolumn]{aastex62}
\usepackage{amsmath}
\usepackage{amssymb}

\newcommand{\dpar}[2]{\frac{\partial#1}{\partial#2}}
\newcommand{\diff}[2]{\frac{\rm{d}#1}{\rm{d} #2}}
\newcommand{\bracketfunc}[3]{\left(\frac{#1}{#2}\right)^{#3}}
\newcommand{\rhosurf}{\Sigma}
\newcommand{\vvec}{\vec{v}}
\newcommand{\vrad}{\rm{v}_{R}}
\newcommand{\vphi}{\rm{v}_{\phi}}
\newcommand{\grav}{\rm{G}}
\newcommand{\mstar}{M_{\ast}}

\newcommand{\mpl}{M_{\rm p}}
\newcommand{\rp}{R_{\rm p}}

\newcommand{\sigmamin}{\rhosurf_{\min}}

\newcommand{\sigmaun}{\rhosurf_{\rm un}}
\newcommand{\sigmaunp}{\rhosurf_{\rm un,p}}
\newcommand{\wgap}{\Delta_{\rm gap}}
\newcommand{\omegakp}{\Omega_{\rm K,p}}
\newcommand{\omegak}{\Omega_{\rm K}}
\newcommand{\tmig}{\tau_{\rm a}}

\newcommand{\rhill}{R_{\rm H}}
\newcommand{\hp}{h_{\rm p}}
\newcommand{\up}{u_{\rm p}}
\newcommand{\uvis}{u_{\rm vis}}
\newcommand{\mdisk}{M_{\rm d}}
\newcommand{\upclassic}{u_{\rm p,cl}}
\newcommand{\Gammaclassic}{\Gamma_{\rm cl}}
\newcommand{\tp}{T_{\rm p}}
\newcommand{\rin}{R_{\rm in}}
\newcommand{\rout}{R_{\rm out}}

\newcommand{\RED}[1]{#1}

\received{}
\revised{}
\accepted{\today}
\submitjournal{ApJ}

\shorttitle{Radial migration of gap-opening planets in protoplanetary disks. I.}
\shortauthors{Kanagawa et al.}

\begin{document}

\title{Radial migration of gap-opening planets in protoplanetary disks. I. The case of a single planet}

\correspondingauthor{Kazuhiro D. Kanagawa}
\email{kazuhiro.kanagawa@usz.edu.pl}

\author[0000-0001-7235-2417]{Kazuhiro D. Kanagawa}
\affiliation{Institute of Physics and CASA$^{\ast}$, Faculty of Mathematics and Physics, University of Szezecin, Wielkopolska 15, PL-70-451 Szczecin, Poland}
\affiliation{Research Center for the Early Universe, Graduate School of Science, University of Tokyo, Hongo, Bunkyo-ku, Tokyo 113-0033, Japan}

\author{Hidekazu Tanaka}
\affiliation{Astronomical Institute, Tohoku University, Sendai, Miyagi 980-8578, Japan}

\author{Ewa Szuszkiewicz}
\affiliation{Institute of Physics and CASA$^{\ast}$, Faculty of Mathematics and Physics, University of Szezecin, Wielkopolska 15, PL-70-451 Szczecin, Poland}

%



\begin{abstract}
A large planet orbiting a star in a protoplanetary disk opens a density gap along its orbit due to the strong disk--planet interaction and migrates with the gap in the disk.
It is expected that in the ideal case, a gap-opening planet migrates at the viscous drift speed, which is referred to as type~II migration.
However, recent hydrodynamic simulations have shown that in general, the gap-opening planet is not locked to the viscous disk evolution.
A new physical model is required to explain the migration speed of such a planet.
For this reason, we re-examined the migration of a planet in the disk, by carrying out the two-dimensional hydrodynamic simulations in a wide parameter range.
We have found that the torque exerted on the gap-opening planet depends on the surface density at the bottom of the gap.
The planet migration slows down as the surface density of the bottom of the gap decreases.
Using the gap model developed in our previous studies, we have constructed an empirical formula of the migration speed of the gap-opening planets, which is consistent with the results given by the hydrodynamic simulations performed by us and other researchers.
Our model easily explains why the migration speed of the gap-opening planets can be faster than the viscous gas drift speed.
It can also predict the planet mass at which the type~I migration is no longer adequate due to the gap development in the disk, providing a gap formation criterion based on planetary migration.
\end{abstract}


\section{Introduction} \label{sec:introduction}
Planets are formed in a protoplanetary disk and migrate in the disk due to gravitational interaction with the surrounding disk gas \citep[e.g.,][]{Lin_Papaloizou1979,Goldreich_Tremaine1980}.
When a planet is small enough that the disk--planet interaction is in the linear regime, the planet migration is referred to as type~I migration, which has been studied in detail both for isothermal \citep[e.g.,][]{Tanaka_Takeuchi_Ward2002} and for non-isothermal disks \citep[e.g.,][]{Paardekooper_Baruteau_Crida_Kley2010}.

When the planet is sufficiently large, the planet forms a density gap along with its orbit due to the strong disk--planet interaction \citep[e.g.,][]{Lin_Papaloizou1993,Ward1997,Lubow_Seibert_Artymowicz1999,Nelson_Papaloizou_Masset_Kley2000,Crida_Morbidelli2007,Edgar2007,Zhu2011,Baruteau_Papaloizou2013,Duffell_Haiman_MacFadyen_DOrazio_Farris2014,Kanagawa2015a,Durmann_Kley2015,Dong_Dawson2016,Dong_Fung2017}.
If the gap is clean and stationary, the migration of the gap-opening planet is referred to as the type~II migration, in which the planet is locked within the gap.
In this picture, the migration speed of the planet corresponds to the speed of the gas viscous accretion when the planet mass ($\mpl$) is much smaller than a local disk mass ($\sigmaunp \rp^2$, where $\sigmaunp$ is a gas surface density of the unperturbed disk at the planet's orbital radius ($\rp$)), which is the so-called disk-dominated case  \citep[e.g.,][]{Lin_Papaloizou1986b,Armitage2007}.
When the planet mass is larger than the local disk mass ($\mpl \gg \sigmaunp \rp^2$), the migration speed of the planet can be slower than the speed of the viscous gas accretion  (the so-called planet-dominated case) \citep[e.g.,][]{Syer_Clarke1995,Ivanov_Papaloizou_Polnarev1999,Armitage2007,Ragusa_Rosotti_Teyssandier_Booth_Clarke_Lodato2018}.
However, in general, the migration of the gap-opening planet is not well understood.
Recent hydrodynamic simulations of the migration of the gap-opening planet \citep{Duffell_Haiman_MacFadyen_DOrazio_Farris2014,Durmann_Kley2015} put more light on this problem.
They have shown that gas can easily cross the gap and the planet is not locked to the viscous disk evolution.
Moreover, these hydrodynamic simulations have demonstrated that the migration speed of the gap-opening planet can be faster than the speed of the viscous gas accretion.
The migration speed faster than the speed of the viscous gas accretion cannot be treated in the 'classical' picture in which the planet is locked into the gap, neither in the planet-dominated case nor the disk-dominated case.
In order to understand the migration of the gap-opening planet, a new physical model is required.

The gap structure plays an important role in determining the migration speed of the gap-opening planet, because the planet moves according to the gravitational torque exerted by the surrounding disk gas.
The gap structure has been investigated in detail by recent hydrodynamic simulations \citep[e.g.,][]{Duffell_MacFadyen2013,Fung_Shi_Chiang2014,Kanagawa2015a,Kanagawa2016a,Kanagawa2017b}.
They have shown that for the usual parameters of the disk, the gap is not very deep even if the planet is as massive as Jupiter.
The detail of this gap model is described in the following section.

The transition between the different type of migrations has been studied in terms of the analytical model \citep{Ward1997} and hydrodynamic simulations \cite[e.g.,][]{Masset_Papaloizou2003,Bate_Lubow_Ogilvie_Miller2003}.
However, a systematic study of the entire parameter space has not been performed till now.
For this reason, we re-examine the migration of the single planet in the protoplanetary disk and construct the new model for the migration speed of the gap-opening planet.
It will be a useful tool for the population synthesis calculations, because the outcome of these calculations depends strongly on the way in which the planets actually migrate \citep[e.g.,][]{Mordasini_Alibert_Benz_Klahr_henning2012,Ida_Lin_Nagasawa2013}.

In this paper, we determine the migration speed of the planet for various planet masses, disk aspect ratios, and viscosities, using the two-dimensional hydrodynamic simulations.
\RED{In Section~\ref{sec:analytical_framework}, before presenting the results of our hydrodynamic simulations, we briefly summarize the classical picture of the type~II migration which does not allow the gas flow across the gap.
Moreover in the same section, we describe the model of the migration if the gas can pass through the gap, which is consistent with the results of recent hydrodynamic simulations.}
We outline basic equations and numerical method in Section~\ref{sec:method}.
In Section~\ref{sec:evolution_one_planet}, we show the results of our hydrodynamic simulations.
We have found that the reduction of the torque exerted on the gap-opening planet is well explained by the decrease in the surface density at the bottom of the gap.
In Section~\ref{sec:empirical_model}, we present the empirical formula of the migration timescale of the planet.
We discuss the effects of corotation torques on the migration of the gap-opening planet in Section~\ref{sec:corotations}.
Section~\ref{sec:summary} includes our summary.

\RED{
\section{Analytical frameworks of the migration of the gap-opening planet} \label{sec:analytical_framework}
\subsection{Classical picture of the type~II migration} \label{subsec:classical_typeII}
Before presenting our hydrodynamic simulations, we briefly summarize the classical picture of the type~II migration and outcomes of this picture.
It is assumed that the planetary gravitational torque is concentrated in a narrow region along with the planet's orbit and the planet forms a deep gap.
Because of the deep gap, a radial gas flow across the gap is halted.
In this case, the planet migrates by being pushed by the gas dammed by the gap, and this gas moves at the same velocity as the planet.
Under above assumptions, the torque exerted on the planet ($\Gammaclassic$) can be calculated as 
\begin{align}
\Gammaclassic &= -3\pi \rp^2 \nu \omegakp \sigmaunp -2\pi \rp^3 \omegakp \sigmaunp \upclassic,
\label{eq:conserv_am_classical_typeII2}
\end{align}
where $R,\omegak,\nu,\Gamma$ are the radial distance from the central star, the Keplerian angular velocity, the kinetic viscosity, and the gravitational torque exerted on the planet, respectively.
The orbital radius of the planet is $\rp$, and in the following, the subscript 'p' indicates the value at $R=\rp$.
The subscript 'cl' denotes the value estimated from the classical picture of the type~II migration.
The migration speed of the planet ($\up$) is related to the torque exerted on the planet as
\begin{align}
\up &= \diff{\rp}{t}=\frac{2\Gamma}{\rp \omegakp \mpl}.
\label{eq:migration_rate}
\end{align}
Equation~(\ref{eq:conserv_am_classical_typeII2}) is exactly the same as Equation~(42) of \cite{Ivanov_Papaloizou_Polnarev1999}.
Solving Equation~(\ref{eq:conserv_am_classical_typeII2}) for $\up$, we obtain
\begin{align}
\upclassic &= \frac{\mdisk}{\mdisk+\mpl} \uvis,
\label{eq:up_classical_typeII}
\end{align}
where $\mdisk=4\pi \rp^2 \sigmaunp$ and $\uvis$ is the radial velocity of viscous gas accretion at $R=\rp$ is written as,
\begin{align}
\uvis &= -\frac{3}{2}\frac{\nu_{\rm p}}{\rp}.
\label{eq:uvis}
\end{align}
We also obtain the torque exerted on the planet in this case, as
\begin{align}
\Gammaclassic &= - \frac{3}{4} \alpha \bracketfunc{\hp}{\rp}{2} \rp^2 \omegakp^2 \frac{\mpl \mdisk}{\mdisk+\mpl},
\label{eq:gamma_classical_typeII}
\end{align}
where we use the $\alpha$-prescription provided by \cite{Shakura_Sunyaev1973}, and hence the kinetic viscosity $\nu$ is expressed by $\alpha (\hp/\rp)^2 R^2 \omegak$, where $\hp$ is the disk scale height at $R=\rp$.

When $\mpl \gg \mdisk$ (the planet-dominated case), the angular momentum flux of gas near the planet is negligible as compared with the planetary torque.
Hence, the torque exerted on the planet is determined by $\Gammaclassic \sim -3\pi \rp^2\nu \omegakp \sigmaunp$.
In this case, the migration speed of the planet is slower than $\uvis$ by a factor of $\mdisk/\mpl$ \citep[e.g.,][]{Syer_Clarke1995,Ivanov_Papaloizou_Polnarev1999,Armitage2007}.
When the mass of the planet is smaller than $\mdisk$ (the disk-dominated case), the angular momentum flux of gas near the planet is not negligible.
In this case, the specific torque of the planet ($\Gamma/\mpl$) and the specific angular momentum flux of the gas ($2\pi \rp^3 \omegak \sigmaunp \up/\mdisk$) are comparable to each other, and the migration speed of the planet corresponds to $\uvis$ \citep[e.g.,][]{Lin_Papaloizou1986b,Armitage2007}.

In the planet-dominated case ($\mpl \gg \mdisk$), the classical picture of the type~II migration predicts the pile-up of the gas at the outer disk, since $\up/\uvis < 1$ and the gas cannot pass through the gap \cite[e.g.,][]{Syer_Clarke1995,Ivanov_Papaloizou_Polnarev1999}.
In this case, it is expected that the structure of the outer disk does not reach steady state and the pile-up of the gas slightly changes the migration speed of the planet from that expected by Equation~(\ref{eq:up_classical_typeII}).
However, the deviation is not significant and hence, one can roughly consider that the migration speed is given by Equation~(\ref{eq:up_classical_typeII}).
Note that recent hydrodynamic simulations with a fixed planet (in which the planetary migration is neglected) \citep[e.g.,][]{Fung_Shi_Chiang2014,Duffell_Haiman_MacFadyen_DOrazio_Farris2014,Durmann_Kley2015,Kanagawa2017b} have succeeded in obtaining the gap structure which hardly changes in time, and the pile-up of the gas in the outer disk has not been observed, even when the calculations have been performed for sufficiently long time.

\subsection{Planetary migration with the radial gas flow across the gap} \label{subsec:picture_with_gaslfow_across_gap}
Recent high-resolution hydrodynamic simulations \citep[e.g.,][]{Duffell_Haiman_MacFadyen_DOrazio_Farris2014,Durmann_Kley2015} have shown that even in the case of the planet as massive as Jupiter,  the gas is able to pass through the gap.
Moreover, these simulations have also demonstrated that when the disk mass is relatively large, the migration speed of the planet can be faster than the speed of the viscous gas accretion.
In this case, Equation~(\ref{eq:conserv_am_classical_typeII2}) is no longer valid and the migration speed of the planet is determined by the sum of the torque exerted on the planet from the inner and outer disks, as
\begin{align}
\Gamma &= -\int^{\rout}_{\rin} R dR \int^{2\pi}_0 d\phi \Sigma \dpar{\Psi}{\phi}  = \int^{\rout}_{\rin} \diff{\Gamma}{R} dR,
\label{eq:gamma}
\end{align}
where $\rin$ and $\rout$ denote the radii of the inner and outer edges of the disk, respectively, and $d\Gamma/dR=- R \int^{2\pi}_{0} (\Sigma \partial \Psi/\partial \phi) d\phi$ is the density of the torque exerted on the planet by the disk, and $\Psi$ and $\Sigma$ are the gravitational potential and the gas surface density, respectively.

The torque from the outer disk (or the inner disk) is related to the gap depth.
In what follows we consider just the outer part of the disk.
As shown by the recent studies \citep{Fung_Shi_Chiang2014,Kanagawa2015b}, the gap depth is determined by the balance between the viscous angular momentum flux and the torque from the outer disk, as
\begin{align}
  3\pi \rp^2 \sigmamin \nu \omegakp - \tp(\rp) &= 3\pi \rp^2 \nu \omegakp \sigmaunp,
\label{eq:conserv_am2}
\end{align}
where $\sigmamin$ is the surface density of the bottom of the gap and $\tp(R)$ is a cumulative torque exerted on the planet by the disk from the radii of $R$ to $\rout$, which is defined as
\begin{align}
\tp(R) &=  \int^{R_{\rm out}}_{R} \diff{\Gamma}{R'} dR'.
\label{eq:cumulative_torque}
\end{align}
Note that $\Gamma = \tp(\rin)$.
We stress that Equation~(\ref{eq:conserv_am2}) considers the radial flow of the gas across the gap (the term of the mass flux is canceled out, since it is constant throughout the disk in steady state condition, see \cite{Kanagawa2015b} for detail derivation).
We also stress that the specific torque exerted on the planet does not need to be similar to that of the disk gas, in this case.
The torque can increase with a larger surface density of the disk gas.
If the gap is wide enough and the planet mainly interacts with the gas at the bottom of the gap (in which $\Sigma \sim \sigmamin$), the torque from the outer disk can be estimated by
\begin{align}
\tp(\rp) \simeq -0.4 \bracketfunc{\mpl}{\mstar}{2} \bracketfunc{\hp}{\rp}{-3} \rp^{4} \omegakp^2 \sigmamin.
\label{eq:one-side_torque}
\end{align}
Substituting Equation~(\ref{eq:one-side_torque}) into Equation~(\ref{eq:conserv_am2}), we obtain $\sigmamin$ as
\begin{align}
\frac{\sigmamin}{\sigmaunp} &= \frac{1}{1+0.04K},
\label{eq:smin}
\end{align}
where
\begin{align}
K&=\bracketfunc{\mpl}{\mstar}{2} \bracketfunc{\hp}{\rp}{-5} \alpha^{-1}.
\label{eq:kgap}
\end{align}
Equation~(\ref{eq:smin}) can reproduce well the gap depth given by the hydrodynamic simulations done by the previous studies \citep{Varniere_Quillen_Frank2004,Duffell_MacFadyen2013,Fung_Shi_Chiang2014,Kanagawa2015b,Fung_Chiang2016,Kanagawa2017b}.
\cite{Kanagawa2017b} have confirmed that the assumption that the planet mainly interacts with the gas at the bottom of the gap leads to the results remaining in a reasonable agreement with the hydrodynamic simulations.
Although the planet migration was not included in the previous studies, we confirm that Equation~(\ref{eq:smin}) is also valid in our hydrodynamic simulations with a migrating planet (see Appendix~\ref{sec:gap_structures}).

The above model of \cite{Kanagawa2015b} (Equation~(\ref{eq:one-side_torque})) cannot predict the torque exerted on the planet, determined by the difference between the cumulative torques exerted on the planet by the inner and outer disks, because it does not consider the asymmetry of the inner and outer disk.
However, if the value of $|\tp(R_{\rm in})/\tp(\rp)|$ is of the order of $\hp/\rp $ as in the linear regime, the torque may be expressed by $\Gamma \sim -(\mpl/\mstar)^2 (\hp/\rp)^{-2} \rp^{4} \omegakp^2 \sigmamin$.
As shown in Section~\ref{sec:evolution_one_planet}, this expectation agrees reasonably well with the results of the hydrodynamic simulations.
}

\section{Basic equations and numerical method} \label{sec:method}
\subsection{Basic equations} \label{subsec:basic_eq}
In our simulations, we use a geometrically thin and non-self-gravitating disk.
We choose a two-dimensional cylindrical coordinate system $(R,\phi)$, and its origin locates at the position of the central star.
The velocity is denoted as $\vvec=(\vrad,\vphi)$, where $\vrad$ and $\vphi$ are the velocities in the radial and azimuthal directions.
The angular velocity is denoted by $\Omega=\vphi/R$.
We adopt a simple isothermal equation of state, in which the vertically integrated pressure $P$ is given by $c_s^2 \rhosurf$, where $c_s$ is the isothermal speed of sound.

The vertically integrated equation of continuity is
\begin{align}
	\dpar{\rhosurf}{t}+\nabla \cdot \left( \rhosurf \vvec \right) &=0.
	\label{eq:mass_eq}
\end{align}
The equations of motion are
\begin{align}
\dpar{\vvec}{t} + \vvec \cdot \nabla \vvec &= \frac{\nabla P}{\rhosurf} - \nabla \Psi + \vec{f}_{\nu},
\label{eq:motion_eq}
\end{align}
where $\vec{f}_{\nu}$ represents the viscous force per unit mass \citep[c.f.,][]{Nelson_Papaloizou_Masset_Kley2000}.
The gravitational potential $\Psi$ is given by the sum of the gravitational potentials of the star and the planet as
\begin{align}
\Psi &= -\frac{\grav \mstar}{R} + \Psi_p + \frac{\grav \mpl}{\rp^2} R\cos\left(\phi-\phi_p \right),
\label{eq:gravpot}
\end{align}
where $\grav$ is the gravitational constant.
The first term of Equation~(\ref{eq:gravpot}) is the potential of the star and the third term represents the indirect terms due to planet--star gravitational interaction.
The second term is the gravitational potential of the planet, which is given by
\begin{align}
\Psi_{p} &= \frac{\grav \mpl}{\left[ R^2+2R\rp \cos\left( \phi-\phi_{p} \right) + \rp^2 +\epsilon^2  \right]^{1/2}},
\label{eq:gravpot_planet}
\end{align}
where $\epsilon$ is a softening parameter.

\subsection{Numerical method} \label{subsec:numerical_method}
To numerically solve Equations~(\ref{eq:mass_eq}) and (\ref{eq:motion_eq}), we use the two-dimensional numerical hydrodynamic code {\sc \tt FARGO}\footnote{See: \url{http://fargo.in2p3.fr/}}, which is an Eulerian polar grid code with a staggered mesh.
Because of a fast advection algorithm that removes the azimuthally averaged velocity for the Courant time step \citep{Masset2000}, we can use {\sc \tt FARGO} to calculate the disk--planet interaction for a long period of time.

The softening parameter $\epsilon$ in the gravitational potential of Equation~(\ref{eq:gravpot_planet}) is set to be $0.6$ times the disk scale height at the location of the planet.
Considering the existence of the circumplanetary disk, we exclude $60\%$ of the planets' Hill radius when calculating the force exerted by the disk on the planet, following \cite{Baruteau_Papaloizou2013}.
For simplicity, we neglect the disk gas accretion onto the planet.

We use unit of the radius as an arbitrary value $R_0$ and unit of the mass as $\mstar$ (the mass of the central star).
\RED{The orbital radius of the planet is initially set to be $R=R_0$.}
The surface density is thus normalized by $\mstar/R_0^2$.
We assume that a power-law distribution of the disk aspect ratio is given by $\hp/\rp = H_0(R/R_0)^f$, where $H_0$ is the disk aspect ratio at $R=R_0$ and $f$ is the flaring index.
We adopt a constant $\alpha$ throughout the disk.
With the flaring index, the temperature varies as $R^{-\beta}$ where $\beta=-2f+1$ and the kinetic viscosity $\nu$ varies as $R^{2f+1/2}$ for a constant $\alpha$.
The initial surface density distribution is given by $\sigmaun(R) = \Sigma_0 (R/R_0)^{-s}$, where $\Sigma_0$ is the initial surface density at $R=R_0$.
We always set $s=2f+1/2$, assuming the steady viscous accretion disk with $\Sigma \propto 1/\nu$.
The initial angular velocity is given by $\omegak \sqrt{1-\eta}$, where $\eta = (1/2)(\hp/\rp)^2 d\ln P/d\ln R$.
The radial drift velocity is given by $v_R = -3\nu/(2R)$.

We adopt the so-called 'open' boundary condition in the inner boundary.
\RED{In the outer boundary, we let the physical quantities (that is, $\rhosurf$, $\vrad$, and $\vphi$) keeping the initial values during the whole simulation.}
To avoid an artificial wave reflection, the damping is used in the so-called wave-killing zone near the boundary layers \citep[c.f.,][]{Val-Borro_etal2006}.
The wave-killing zones are located from $R_{\rm out}-0.2R_0$ to $R_{\rm out}$ for the outer boundary and from $R_{\rm in}$ to $R_{\rm in}+0.1R_0$ for the inner boundary, where $R_{\rm out}$ and $R_{\rm in}$ are the radius of the outer and inner boundaries, respectively.

We define $t_0$ as $2\pi/\Omega_0$, where $\Omega_0$ is the Keplerian angular velocity at $R=R_0$.
The characteristic torque on the planet without a gap is approximately given by
\begin{align}
\Gamma_0 (R)&= \bracketfunc{\mpl}{\mstar}{2}\bracketfunc{h}{R}{-2} \sigmaun R^4 \omegak^2.
\label{eq:gamma0}
\end{align}
Note that $\Gamma_0$ depends on $R$ as well as the unperturbed surface density, the aspect ratio and angular velocity.
In the linear theory, the torque exerted on the planet is obtained as the sum of the Lindblad ($\Gamma_L$) and the corotation torques ($\Gamma_C$).
In the locally isothermal case, these torques are given by \citep{Paardekooper_Baruteau_Crida_Kley2010},
\begin{align}
\frac{\Gamma_L}{\Gamma_0(\rp)} &= -\left(2.5-0.1s+1.7 \beta \right) b^{0.71}
\label{eq:GammaL_Paardekooper11}\\
\frac{\Gamma_C}{\Gamma_0(\rp)} &= 1.1\left(1.5-s\right)b +2.2\beta b^{0.71} - 1.4\beta b^{1.26},
\label{eq:GammaC_Paardekooper11}
\end{align}
where $b=(0.4\hp)/\epsilon$, and $T \propto R^{-\beta}$.
When $\epsilon=0.6\hp$, $s=0.5$ and $f=0$, the torque exerted on the planet is given by $-1.6\Gamma_0(\rp)$.

\section{Results of hydrodynamic simulations} \label{sec:evolution_one_planet}
In this section, we present results of numerical hydrodynamic simulations.
We adopt a computational domain from $R_{\rm in} = 0.4R_0$ to $R_{\rm out}=3.0R_0$.
The domain is divided in $512$ meshes in the radial direction (equally spaced in logarithmic space) and in $2048$ meshes in the azimuthal direction (equally spaced).
We explore the migration timescale of the planet by varying the mass of the planet ($5\times10^{-6}<\mpl/\mstar<1\times 10^{-3}$), the disk aspect ratio ($0.03<H_0<0.07$), and viscosity ($10^{-2}<\alpha<10^{-3}$).
The parameter sets of the simulations are listed in Table~\ref{tab:models_oneplanet} (98 runs).
\RED{We adopt $\Sigma_0=1\times 10^{-4}$ in these simulations, which corresponds to $8.9\mbox{ g/cm}^2$ when $\mstar=1M_{\ast}$ and $R_0=10\mbox{ AU}$ since it is scaled by $\mstar/R_0^2$.}

\begin{table}
 \caption{Parameters (98 runs)}
 \label{tab:models_oneplanet}
 \begin{center}
 \begin{tabular}{cccc}
  \hline \hline
  $\mpl/\mstar$ &$H_0$ &$\alpha$ & $f(s=2f+1/2)$\\
  \hline \hline
  $1\times10^{-5}$& $0.05$ &$10^{-2},5\times10^{-3},10^{-3}$ &$0 (1/2)$, $1/4(1)$\\
  $5\times10^{-5}$& $0.05$ &$10^{-2},5\times10^{-3},10^{-3}$ &$0 (1/2)$, $1/4(1)$\\
  $8\times10^{-5}$& $0.05$ &$10^{-2},5\times10^{-3},10^{-3}$ &$0 (1/2)$, $1/4(1)$\\
  $1\times10^{-4}$& $0.05$ &$10^{-2},5\times10^{-3},10^{-3}$ &$0 (1/2)$, $1/4(1)$\\
  $3\times10^{-4}$& $0.05$ &$10^{-2},5\times10^{-3},10^{-3}$ &$0 (1/2)$, $1/4(1)$\\
  $5\times10^{-4}$& $0.05$ &$10^{-2},5\times10^{-3},10^{-3}$ &$0 (1/2)$, $1/4(1)$\\
  $8\times10^{-4}$& $0.05$ &$10^{-2},5\times10^{-3},10^{-3}$ &$0 (1/2)$, $1/4(1)$\\
  $1\times10^{-3}$& $0.05$ &$10^{-2},5\times10^{-3},10^{-3}$ &$0 (1/2)$, $1/4(1)$\\
  \hline
  $5\times10^{-6}$& $0.03$ &$10^{-2},10^{-3}$ &$0 (1/2)$, $1/4(1)$\\
  $1\times10^{-5}$& $0.03$ &$10^{-2},10^{-3}$ &$0 (1/2)$, $1/4(1)$\\
  $3\times10^{-5}$& $0.03$ &$10^{-2},10^{-3}$ &$0 (1/2)$, $1/4(1)$\\
  $5\times10^{-5}$& $0.03$ &$10^{-2},10^{-3}$ &$0 (1/2)$, $1/4(1)$\\
  $8\times10^{-5}$& $0.03$ &$10^{-2},10^{-3}$ &$0 (1/2)$, $1/4(1)$\\
  $1\times10^{-4}$& $0.03$ &$10^{-2},10^{-3}$ &$0 (1/2)$, $1/4(1)$\\
  $3\times10^{-4}$& $0.03$ &$10^{-2},10^{-3}$ &$0 (1/2)$, $1/4(1)$\\
  $5\times10^{-4}$& $0.03$ &$10^{-2},10^{-3}$ &$0 (1/2)$, $1/4(1)$\\
  \hline
  $5\times10^{-5}$& $0.07$ &$10^{-2},10^{-3}$ &$0 (1/2)$, $1/4(1)$\\
  $1\times10^{-4}$& $0.07$ &$10^{-2},10^{-3}$ &$0 (1/2)$, $1/4(1)$\\
  $5\times10^{-4}$& $0.07$ &$10^{-2},10^{-3}$ &$0 (1/2)$, $1/4(1)$\\
  $1\times10^{-3}$& $0.07$ &$10^{-2},10^{-3}$ &$0 (1/2)$, $1/4(1)$\\
  \hline
 \end{tabular}
 \end{center}
\end{table}

\subsection{Time variation of planetary torque} \label{subsec:timevar_torque}
\begin{figure*}
	\begin{center}
		\resizebox{0.98\textwidth}{!}{\includegraphics{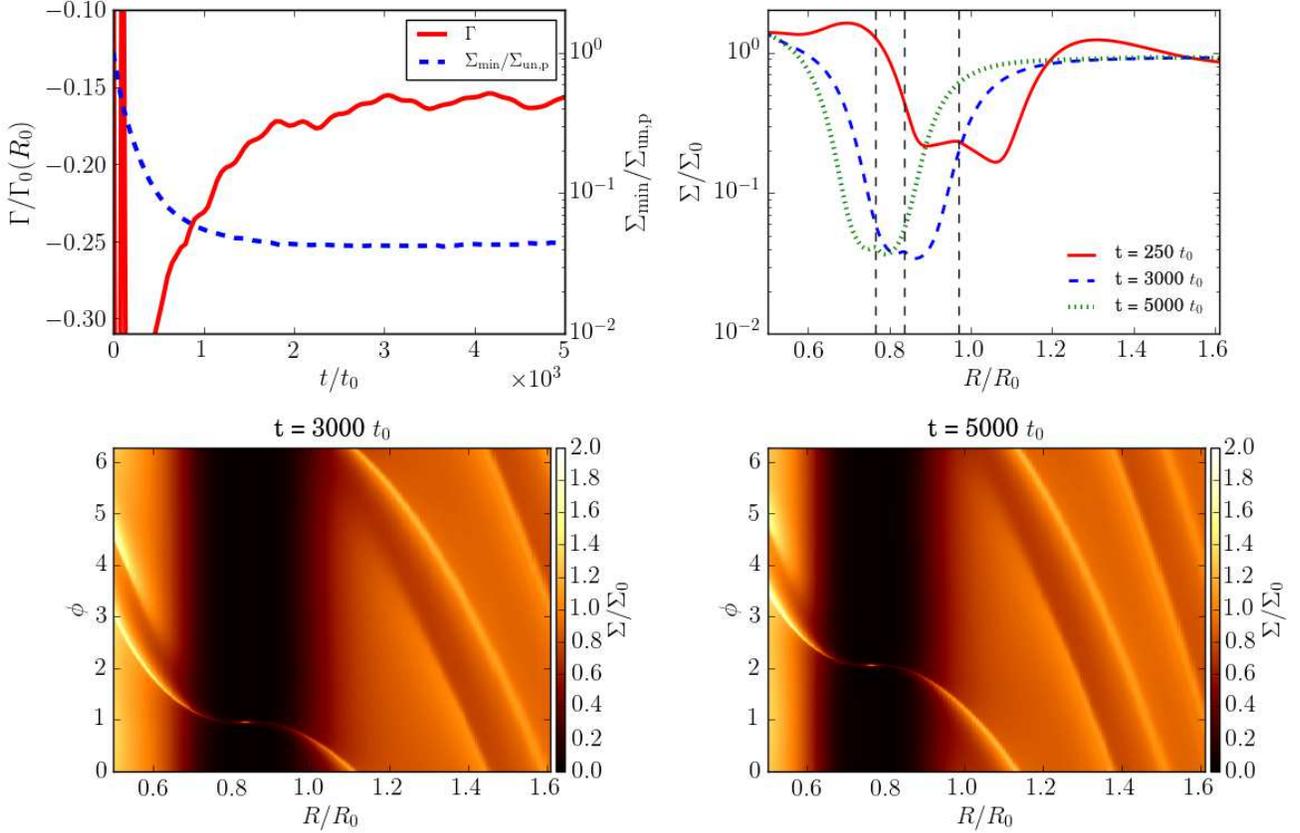}}
		\caption{
		Time variation of the torque exerted on the planet and the surface density at the bottom of the gap (upper left) in the case of $\mpl/\mstar=5\times 10^{-4}$, $H_0=0.05$, and $\alpha=10^{-3}$.
		The flaring index is set to be $0$.
		In the upper right, the distributions of the azimuthally averaged surface density are illustrated.
		The vertical line dashed lines in the upper right panel indicate the position of the planet at the given moment of time.
		In the lower panels, the 2D distributions of the surface density are illustrated.
		The color indicates the surface density of the gas in the lower panels.
		\label{fig:evo_dens_tmig_q5e-4_a1e-3_h0.05}
		}
	\end{center}
\end{figure*}
The upper left panel of Figure~\ref{fig:evo_dens_tmig_q5e-4_a1e-3_h0.05} shows the time variations of the torque exerted on the planet and the surface density at the bottom of the gap ($\sigmamin$) in the case with $\mpl/\mstar=5\times10^{-4}$, $H_0=0.05$, and $\alpha=10^{-3}$.
The flaring index is set to be $0$ (and $s=0.5$) in the figure.
Following \cite{Fung_Shi_Chiang2014}, we define $\sigmamin$ as an average over the annulus spanning from $R=\rp-\delta$ to $R=\rp+\delta$ with $\delta=2\max(\rhill,\hp)$, excised from $\phi = \phi_{\rm p}-\delta/\rp$ to $\phi=\phi_{\rm p}+\delta/\rp$.
In the upper right panel of the figure, we plot the distributions of the azimuthally averaged surface density at $t=250t_0$, $3000t_0$, and $5000t_0$.
In the lower panels of Figure~\ref{fig:evo_dens_tmig_q5e-4_a1e-3_h0.05}, the two dimensional distributions of the surface densities at $t=3000t_0$ and $t=5000t_0$ are illustrated.
The torque exerted on the planet is saturated after $t\simeq 3000 t_0$, and $\sigmamin$ reaches the stationary value around the same time.
As can be seen from the upper right panel of Figure~\ref{fig:evo_dens_tmig_q5e-4_a1e-3_h0.05}, the gap structure (i.e., the depth and the width) is almost unchanged after $t\simeq 3000 t_0$ and its shape is always axisymmetric as can be seen in the lower panels.
\RED{In this case, the gap is relatively deep ($\sigmamin/\sigmaunp \simeq 4\times 10^{-2}$), and the local disk mass is comparable with the planet mass ($\mdisk/\mstar \simeq 10^{-3}$ and $\mpl/\mstar=5\times 10^{-4}$).
Nevertheless, the stationary value of $|\Gamma/\Gamma_0(R_0)|$ is about $0.15$, which is about 1.5 times larger than $|\Gammaclassic/\Gamma_0(R_0)|$ ($\simeq 3\pi \nu \rp^2 \omegakp\mpl/\Gamma_0(R_0)=0.09$ in this case) given by Equation~(\ref{eq:gamma_classical_typeII}).
Such a large torque cannot be explained by the classical picture of the type~II migration.

\begin{figure}
	\begin{center}
 		\resizebox{0.49\textwidth}{!}{\includegraphics{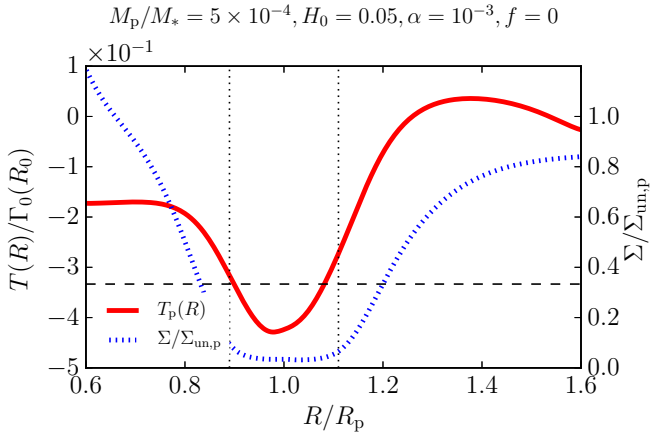}}
		\caption{
 		The distributions of the cumulative torques and the azimuthal averaged surface density at $t=3000t_0$, in the same case as in Figure~\ref{fig:evo_dens_tmig_q5e-4_a1e-3_h0.05}.
 		The vertical dotted lines denote the positions of $\rp \pm 2\mbox{max}(R_H,h_p)$, and the horizontal dashed line denotes $\sigmaunp/3$, respectively.
		\label{fig:cumulative_torque_q5e-4_h0.05_a1e-3}
		}
	\end{center}
\end{figure}
Figure~\ref{fig:cumulative_torque_q5e-4_h0.05_a1e-3} shows the distributions of the cumulative torques (which is defined by Equation~(\ref{eq:cumulative_torque})), for the case presented in Figure~\ref{fig:evo_dens_tmig_q5e-4_a1e-3_h0.05} at $t=3000t_0$.
In this case, the cumulative torques increases from the bottom of the gap (the region with $\Sigma < \sigmamin$) to the edge of the gap (the surface density is a bit larger than $\sigmamin$, but is significantly smaller than the density of the unperturbed disk).
It is reasonable to consider $\sigmamin$ as a representative surface density of the gas with which the planet mainly interacts.
This indicates that the torque exerted on the planet is determined by the interaction with the gas within the gap, rather than the interaction with the gas outside the gap.
This is consistent with the time variations of $\Gamma$ and $\sigmamin$ as shown in the upper left panel of Figure~\ref{fig:evo_dens_tmig_q5e-4_a1e-3_h0.05}.
}

In the runs listed in Table~\ref{tab:models_oneplanet}, the gap depth and torque exerted on the planet approximately reach the stationary value at $t=3000t_0$, as in Figure~\ref{fig:evo_dens_tmig_q5e-4_a1e-3_h0.05}.
In the following, we regard the torque at $t=3000t_0$ as the value in steady state.
Note that in few runs\footnote{Specifically, following three runs with $H_0=0.07$ and $\alpha=10^{-2}$:
$\mpl/\mstar=5\times10^{-4}$ in the case of $f=1/4$ and $\mpl/\mstar=10^{-3}$ in the cases of $f=0$ and $f=1/4$.}, \RED{since the viscous timescale is short and the migration speed is too fast, the planet reaches the inner boundary, before $t=3000t_0$.}
In this case, we regard the torque at $\rp/R_0=0.7$ as the stationary value.

\subsection{Planetary torques in steady states} \label{subsec:torque}
\begin{figure}
	\begin{center}
		\resizebox{0.49\textwidth}{!}{\includegraphics{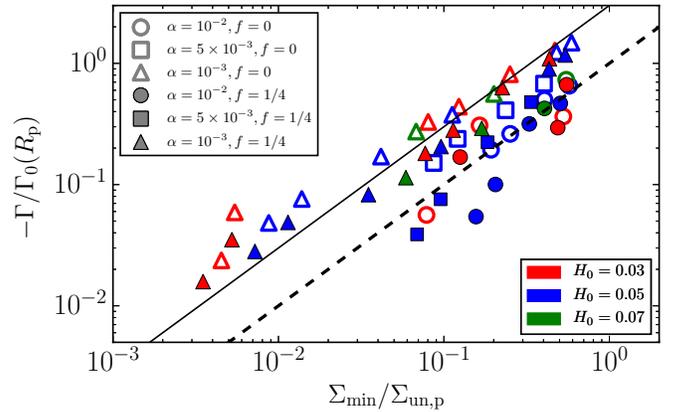}}
		\caption{
		The torque exerted on the planet as a function of the surface density at the bottom of the gap, only in those cases in which $\sigmamin/\sigmaunp<0.6$.
		The solid and dashed lines indicate Equation~(\ref{eq:gamma_vs_sigma}) with $c=3$ and $c=1$, respectively.
		Symbols indicate the values of $\alpha$ (open symbols for $f=0$, and filled symbols for $f=1/4$) while colors indicate values of $H_0$.
		\label{fig:torque_vs_smin}
		}
	\end{center}
\end{figure}
In Figure~\ref{fig:torque_vs_smin}, we show the torque exerted on the planet when the gap is relatively deep.
In the figure, we put on display only those cases in which $\sigmamin$ is smaller than $0.6\sigmaunp$ for clarity purpose.
As can be seen in the figure, the torque is roughly proportional to the surface density at the bottom of the gap and might be expressed as follows,
\begin{align}
\Gamma &= -c \frac{\sigmamin}{\sigmaunp} \Gamma_0(\rp),
\label{eq:gamma_vs_sigma}
\end{align}
where $c$ is a proportionality coefficient.
In the cases of $\alpha=10^{-3}$, the value of $-\Gamma/\Gamma_0(\rp)$ is in good agreement with Equation~(\ref{eq:gamma_vs_sigma}) with $c=3$.
In the cases of $\alpha=10^{-2}$ and $5\times10^{-3}$, the value of $-\Gamma/\Gamma_0(\rp)$ is slightly smaller than that in the case of $\alpha=10^{-3}$ even if $\sigmamin/\sigmaunp$ is similar.
\RED{In this case, Equation~(\ref{eq:gamma_vs_sigma}) with $c=1$ gives a better fit, rather than that with $c=3$.
This correspondence is particularly good when the gap is not very deep.
}
Although the torques in the cases with $f=0$ and $f=1/4$ are slightly different, the dependence of the torque on the flaring index does not seem to be significant.

\RED{
The torque which is roughly proportional to the surface density at the bottom of the gap can be naturally explained by noticing that the gap-opening planet mainly interacts with the gas at the bottom of the gap.
Hence, the decrease in the torque exerted on the planet is simply because of the decrease of the surface density at the bottom of the gap, as for the small planet in the type~I regime  (Equations~(\ref{eq:GammaL_Paardekooper11}) and (\ref{eq:GammaC_Paardekooper11})).
This explanation is also consistent with the fact that $\sigmamin$ can be reasonably well reproduced by Equation~(\ref{eq:smin}), as mentioned in Section~\ref{subsec:picture_with_gaslfow_across_gap}.
If the gap is extremely deep, the gas outside the gap would mainly contribute to the torque exerted on the planet, rather than the gas within the gap, as in the classical picture of the type~II migration.
However, within the parameter range of our survey, the planets do not form such deep gaps.
}

In some cases, especially with a relatively high viscosity (i.e., $\alpha=10^{-2}$ and $\alpha=5\times 10^{-3}$), the corotation torque is positively boosted due to the nonlinear effects, as predicted by \cite{Masset_DAngelo_Kley2006} and \cite{Duffell2015b}.
In consequence, the value of $-\Gamma$ is reduced by the boosted positive corotation torques in these cases.
The scatter of the computational data in Figure~\ref{fig:torque_vs_smin} is partially related to the effects of the nonlinear corotation torque.
We discuss about the effects of the corotation torque in Section~\ref{sec:corotations}.
\RED{
Note that, in two cases ($\mpl/\mstar=3\times 10^{-5}$ and $f=0$, $\mpl/\mstar=5\times 10^{-4}$ and $f=1/4$ with $H_0=0.03$, $\alpha=10^{-2}$), the torque is almost zero or positive, due to the boost of the corotation torque.
For that reason, we do not plot them in Figure~\ref{fig:torque_vs_smin}.
These cases are shown in Section~\ref{sec:corotations}.
}

\RED{
One can find out that the scatter of the computational data in Figure~\ref{fig:torque_vs_smin} is related to the value of $\alpha$, namely the absolute value of the torque is smaller for larger value of $\alpha$ even when $\sigmamin$ is similar.
As mentioned above, this dependence is partially connected to the corotation torque.
In this paper, we do not give any corrections of our formula for the effects of the nonlinear corotation torque, since the nonlinear effects are still not clearly understood.
Further deep investigations about the nonlinear effects of the corotation torque are required to obtain more accurate formula.
}

\begin{figure}
	\begin{center}
		\resizebox{0.49\textwidth}{!}{\includegraphics{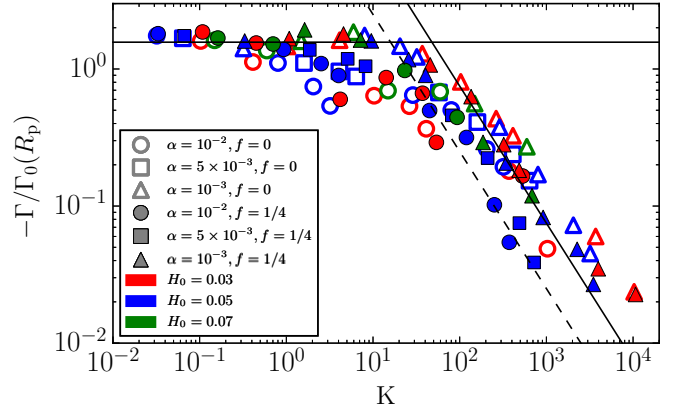}}
		\caption{
		The torque exerted on the planet as a function of $K$.
		The horizontal solid line shows the torque expected by the linear theory in Equations~(\ref{eq:GammaL_Paardekooper11}) and (\ref{eq:GammaC_Paardekooper11}) ($s=0.5$ and $\epsilon=0.6\hp$).
		The solid and dashed declined lines show Equation~(\ref{eq:gamma_vs_kgap}) with $c=3$ and $c=1$, respectively.
		Marks indicate values of $\alpha$ (open marks for $f=0$, and filled marks for $f=1/4$) while colors indicate values of $H_0$.
		\label{fig:torque_vs_kgap}
		}
	\end{center}
\end{figure}

Combining Equations~(\ref{eq:gamma_vs_sigma}) and (\ref{eq:smin}), we obtain the expression for the torque as a function of $K$,
\begin{align}
\Gamma &= - \frac{c}{1+0.04K} \Gamma_0(\rp).
\label{eq:gamma_vs_kgap}
\end{align}
For every run listed in Table~\ref{tab:models_oneplanet} (with the exceptions of two runs of $\mpl/\mstar=3\times 10^{-5}$ with $f=0$ and $\mpl/\mstar=5\times 10^{-4}$ with $f=1/4$ with $H_0=0.03$ and $\alpha=10^{-2}$), we plot the relation between $-\Gamma/\Gamma_0(\rp)$ and $K$ in Figure~\ref{fig:torque_vs_kgap}.
The torque provided by the simulations for small value of K are consistent with the type~I torque given by the sum of Equations~(\ref{eq:GammaL_Paardekooper11}) and (\ref{eq:GammaC_Paardekooper11}), represented by the solid horizontal line in Figure~\ref{fig:torque_vs_kgap}.
For large values of $K$ namely $0.04K\gg 1$ ($K\gg 25$) (when the gap is relatively deep), we can rewrite Equation~(\ref{eq:gamma_vs_kgap}) as
\begin{align}
\Gamma &= -\frac{c}{0.04K}\Gamma_0(\rp) \nonumber\\
&=-25c \alpha \bracketfunc{\hp}{\rp}{3} \sigmaunp \rp^4 \omegakp^2.
\label{eq:gamma_vs_kgap_deepgap}
\end{align}
As can be seen in Figure~\ref{fig:torque_vs_kgap}, when $K\gtrsim 100$, $-\Gamma/\Gamma_0$ given by the simulations are inversely proportional to $K$, in agreement with Equation~(\ref{eq:gamma_vs_kgap_deepgap}).
When $\alpha=10^{-3}$, the calculated torques are consistent with Equation~(\ref{eq:gamma_vs_kgap_deepgap}) with $c=3$.
When $\alpha=10^{-2}$ and $\alpha=5\times 10^{-3}$, Equation~(\ref{eq:gamma_vs_kgap_deepgap}) with $c=1$ gives a better fit than that with $c=3$.

\cite{Duffell_Haiman_MacFadyen_DOrazio_Farris2014} have obtained the empirical formula for the migration speed of the Jupiter-sized planet.
When $\sigmaunp \rp^2/\mpl \ll 1$, their formula gives $\Gamma$ in the form,
\begin{align}
\Gamma &= -\frac{\omega_0}{2} \sigmaunp \rp^4 \omegakp^2.
\label{eq:gamma_d14}
\end{align}
The fitting parameter $\omega_0$ is given by $2.8\times 10^{-4}$ for $\hp/\rp=0.05$ ($\alpha=10^{-2}$)\footnote{\cite{Duffell_Haiman_MacFadyen_DOrazio_Farris2014} adopt $\nu=2.5\times 10^{-5}$, which is constant throughout the disk.
Hence, the value of $\alpha$ changes with the disk aspect ratio.} and by $6.4\times 10^{-5}$ for $\hp/\rp=0.025$ ($\alpha=4\times 10^{-2}$).
Comparing Equations~(\ref{eq:gamma_vs_kgap_deepgap}) and (\ref{eq:gamma_d14}), we obtain $\omega_0=50c\alpha(\hp/\rp)^3$ which is given by $1.8\times 10^{-4}(c/3)$ if $\hp/\rp=0.05$ and by $9.4\times 10^{-5} (c/3)$ if $\hp/\rp=0.025$.
Hence Equation~(\ref{eq:gamma_vs_kgap_deepgap}) with $c=3$ reasonably agrees with the formula given by \cite{Duffell_Haiman_MacFadyen_DOrazio_Farris2014}.
\RED{
In the classical picture of the type~II migration, on the other hand, Equation~(\ref{eq:gamma_classical_typeII}) predict $\omega_0 = 6\pi \alpha (\hp/\rp)^2  = 4.71\times 10^{-4}$, which is larger than that given by \cite{Duffell_Haiman_MacFadyen_DOrazio_Farris2014}.}

When the planet forms the relatively deep gap, the torque is inversely proportional to the value of $K$ as shown above.
When the planet is small enough, on the other hand, the migration speed is described by type~I migration (linear theory), which is independent of the value of $K$.
Around $K\sim 20$, we can find the transition form the planet migrating in the type~I migration to the migration of the gap-opening planet.
When $K\sim 20$, the surface density at the bottom of the gap is $\simeq 0.5\sigmaunp$ (see Equation~(\ref{eq:smin})) and the planet mass is given by
\begin{align}
\bracketfunc{\mpl}{\mstar}{}_{\rm trans} &= 8\times 10^{-5} \bracketfunc{\alpha}{10^{-3}}{1/2} \bracketfunc{\hp/\rp}{0.05}{5/2}.
\label{eq:qtrans}
\end{align}
When the planet mass is larger than $(\mpl/\mstar)_{\rm trans}$, the torque reduces as the surface density at the bottom of the gap decreases, and then the migration speed of the planet is slow-down due to the gap formation.
In terms of the planetary migration, we can consider the condition of $K\gtrsim 20$ as the gap formation criterion.

A gap formation criterion has been considered by many previous works \citep[e.g.,][]{Lin_Papaloizou1993,Rafikov2002b,Crida_Morbidelli_Masset2006,Edgar_Quillen_Park2007}.
For example, \cite{Crida_Morbidelli_Masset2006} have provided the gap formation condition which corresponds to the requirement that the surface density at the bottom of the gap is smaller than $10\%$ of the unperturbed value (Equation~(15) of that paper).
\RED{
As can be seen in Figure~\ref{fig:torque_vs_smin}, for the planetary migration, the torque is affected by the gap formation and decreases along with decreasing $\sigmamin$, even if the surface density at the bottom of the gap is larger than $10\%$ of the unperturbed surface density.
Because of it, Equation~(\ref{eq:qtrans}) is different from that given by Crida's criterion.
}

\RED{
When the viscosity is very small, there is a minimum mass of the planet which can form the gap.
The minimum mass of the planet for the gap formation is given by \cite{Rafikov2002b} as $\mpl \lesssim 2 M_{\oplus}$ when $\alpha \lesssim 10^{-4}$.
Equation~(\ref{eq:qtrans}) is not valid if the planet mass is smaller than this minimum mass of the planet for the gap formation.
In addition, in the case of an extremely small viscosity ($\alpha \ll 10^{-4}$), even small planet may be able to form multiple gaps \citep{Bae_Zhu_Hartmann2017} or a very deep and wide gap \citep{Ginzburg_Sari2018}.
The Rossby wave instability also should be considered in the case of an extremely small viscosity, since it must influence the gap structure \citep[e.g.,][]{Li_etal2000,Lin2014,Ono2016}.
In this case, the planetary migration would be different from that we consider in this paper.
}

\subsection{Additional runs for massive disks} \label{subsec:addition_runs}
\begin{table}
	\begin{center}
 		\caption{Runs with initial gaps (19 runs) ($f=0$)}
 		\label{tab:models_initial_gaps}
 		\begin{tabular}{cc|cc}
 		\hline \hline
 		\multicolumn{2}{c|}{$H_0=0.05$}&\multicolumn{2}{c}{$H_0=0.03$}\\
 		\multicolumn{2}{c|}{$(\mpl/\mstar=10^{-3})$}&\multicolumn{2}{c}{$(\mpl/\mstar=5\times 10^{-4})$}\\
 		\hline
  		$\Sigma_0$ &$\alpha$ & $\Sigma_0$ & $\alpha$\\
  		\hline
  		$5\times 10^{-5}$ & $10^{-2},10^{-3}$ & $1\times 10^{-5}$ & $10^{-2},10^{-3}$\\
  		$1\times 10^{-4}$ & $10^{-2},10^{-3}$ & $5\times 10^{-5}$ & $10^{-2},10^{-3}$\\
  		$5\times 10^{-4}$ & $10^{-2},10^{-3}$ & $1\times 10^{-4}$ & $10^{-2},10^{-3}$\\
  		$1\times 10^{-3}$ & $10^{-3}$         & $5\times 10^{-4}$ & $10^{-2},10^{-3}$\\
  		$3\times 10^{-3}$ & $10^{-3}$         & $1\times 10^{-3}$ & $10^{-2},10^{-3}$\\
  		                  &                   & $3\times 10^{-3}$ & $10^{-3}$\\
  		\hline
 		\end{tabular}
 	\end{center}
\end{table}
As shown by the previous works \citep[e.g.,][]{Edgar2007,Duffell_Haiman_MacFadyen_DOrazio_Farris2014,Durmann_Kley2015}, the torque exerted on the planet depends on the unperturbed disk surface density ($\sigmaunp$).
In the simulations listed in Table~\ref{tab:models_oneplanet}, we do not change the value of $\sigmaunp$.
In order to examine the dependence on the value of $\sigmaunp$, we perform additional simulations for various values of $\sigmaunp$.

When $\sigmaunp$ is large, the migration speed is very fast and the planets reaches the inner edge of the computational domain before the gap shape reaches its steady state.
To avoid this fast migration at the beginning, we adopt the gap structure developed by \cite{Kanagawa2017b} (Equation~(6) of that paper) as an initial surface density distribution.
The setup of the simulations is the same as those described in Section~\ref{subsec:numerical_method}, except the initial distribution of the surface density.
Varying the value of $\Sigma_0$, we examine the migration of the Jupiter-sized planet ($\mpl/\mstar=10^{-3}$) in the case of $H_0=0.05$ and that of the planet with $\mpl/\mstar=5\times 10^{-4}$ in the case of $H_0=0.03$.
The parameters are summarized in Table~\ref{tab:models_initial_gaps} (19 runs).

\begin{figure}
	\begin{center}
		\resizebox{0.49\textwidth}{!}{\includegraphics{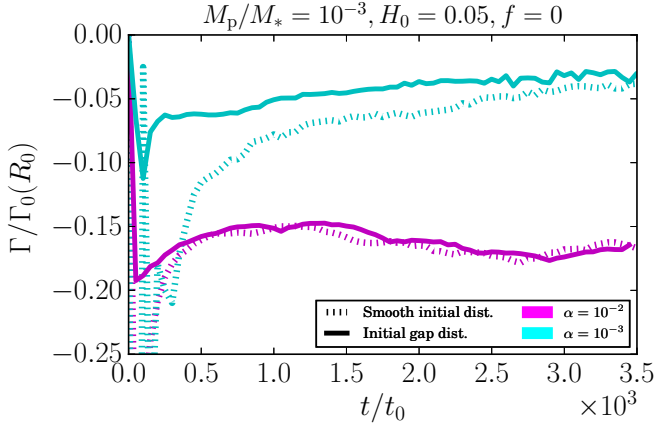}}
		\caption{
		Time variations of the torque exerted on the planet of $\mpl/\mstar=10^{-3}$, in the cases of $\alpha=10^{-2}$ (magenta) and $10^{-3}$ (cyan).
		For the dashed lines, we adopt the smooth surface density distributions (power-law) as the initial condition.
		For the solid lines, the runs start from the surface density with the gap obtained from Equation~(6) of \protect \cite{Kanagawa2017b}.
		The aspect ratio and the flaring index are $0.05$ and $0$, respectively.
		\label{fig:check_convergence_torque}
		}
	\end{center}
\end{figure}
We check the convergence of the torque in the simulations with the initial gap and with the smooth power-law surface density distribution (no gap).
In Figure~\ref{fig:check_convergence_torque}, we compare the time variations of the torques obtained from the simulations which start from the smooth surface density distributions and surface density with the initial gap, when $\mpl/\mstar=10^{-3}$, $H_0=0.05$, in the cases of $\alpha=10^{-2}$ and $\alpha=10^{-3}$.
As can be seen in the figure, in the cases with the initial gap, the torques are saturated quickly.
This saturated torque in the case with the initial gap is consistent to that in the case with the smooth initial surface density distributions.

\subsection{Migration speed of gap-opening planets} \label{subsec:migrate_rate_typeII}
\begin{figure}
	\begin{center}
		\resizebox{0.49\textwidth}{!}{\includegraphics{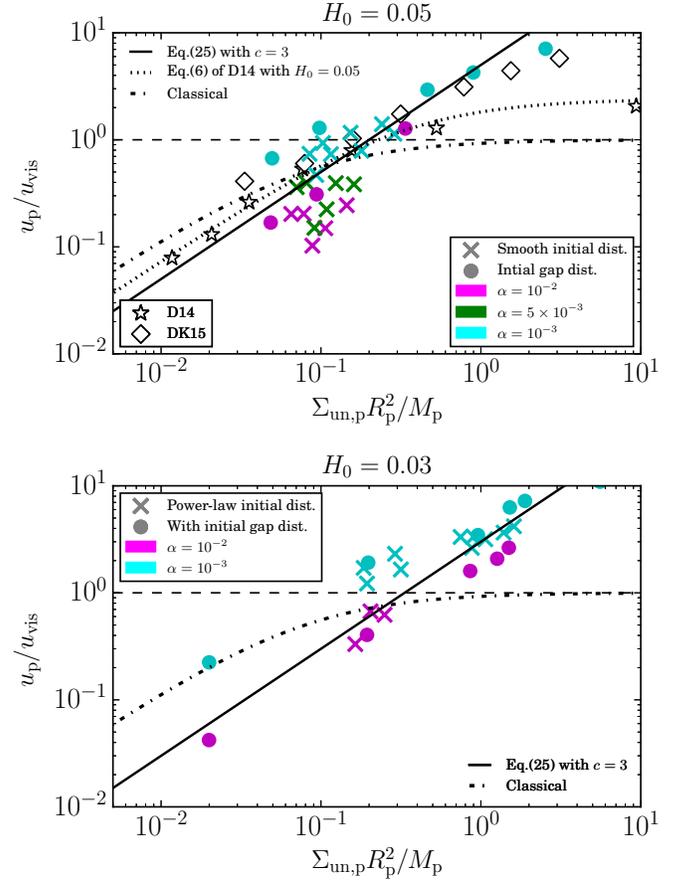}}
		\caption{
		The ratio of $\up$ to $\uvis$ given by the hydrodynamic simulations, for $H_0=0.05$ (upper panel) and $H_0=0.03$ (lower panel), in the cases of $K>100$.
		The crosses and circles indicate the ratios given in the simulations with the smooth initial surface density distribution and the initial gap, respectively.
		The colors indicate the difference values of $\alpha$.
		The solid lines are Equation~(\ref{eq:up_vs_uvis_typeII}) with $c=3$ for $H_0=0.05$ (upper panel) and $H_0=0.03$ (lower panel), respectively.
		In the upper panel,  stars and diamonds indicate the ratios given by Figure~3 of \protect \cite{Duffell_Haiman_MacFadyen_DOrazio_Farris2014} and Figure~15 of \protect \cite{Durmann_Kley2015}, respectively.
		The dot-dashed lines in the upper and lower panels indicate the ratio given by Equation~(\ref{eq:up_classical_typeII}), which is predicted by the classical picture of the type~II migration.
		\label{fig:up_uvis_ratio_vs_diskmass}
		}
	\end{center}
\end{figure}
Here we focus on the migration of the gap-opening planet with $K>100$.
Using Equations~(\ref{eq:migration_rate}) and (\ref{eq:gamma_vs_kgap_deepgap}), we obtain the migration speed ($\up$) for the gap-opening planet as
\begin{align}
\up &= -50c \frac{\Gamma_0}{\rp \omegakp \mpl} \frac{1}{K} \nonumber\\
&= -50c \frac{\sigmaunp \rp}{\mpl} \alpha \bracketfunc{\hp}{\rp}{3} \omegakp.
\label{eq:migrate_typeII}
\end{align}
Then, from Equations~(\ref{eq:migrate_typeII}) and (\ref{eq:uvis}), the ratio of $\up$ to $\uvis$ for the gap-opening planet is given by
\begin{align}
\frac{\up}{\uvis} &= c\frac{\left(\hp/\rp\right)}{0.03} \frac{\sigmaunp \rp^2}{\mpl}.
\label{eq:up_vs_uvis_typeII}
\end{align}
\RED{
In the classical picture of the type~II migration, the ratio of $\up$ to $\uvis$ depends only on the ratio of the local disk mass to the planet mass (Equation~(\ref{eq:up_classical_typeII})), as described in Section~\ref{subsec:classical_typeII}.
On the other hand, Equation~(\ref{eq:up_vs_uvis_typeII}) shows that the ratio $\up/\uvis$ is proportional to the disk aspect ratio, as well as $\sigmaunp \rp^2/\mpl$.
}
We plot the ratio of $\up$ to $\uvis$ obtained from the hydrodynamic simulations with the power-law initial distribution and with the initial gap, in Figure~\ref{fig:up_uvis_ratio_vs_diskmass}.
\RED{
The ratio predicted by the classical picture of the type~II migration (Equation~(\ref{eq:up_classical_typeII})) is also plotted in the figure.
}

In the cases of $H_0=0.05$ (the upper panel of Figure~\ref{fig:up_uvis_ratio_vs_diskmass}), the ratios of $\up$ to $\uvis$ obtained from our simulations are distributed around Equation~(\ref{eq:up_vs_uvis_typeII}) with $c=3$, in both runs with the smooth power-law initial surface density distribution and the initial gap, though being a bit scattered in the cases with the smooth initial distribution.
\cite{Duffell_Haiman_MacFadyen_DOrazio_Farris2014} and \cite{Durmann_Kley2015} have examined the migration speed of Jupiter-sized planet, varying the disk surface density.
We plot the ratios obtained from their simulations in the upper panel of Figure~\ref{fig:up_uvis_ratio_vs_diskmass}.
The ratios given by \cite{Durmann_Kley2015} agree well with the results of our simulations and Equation~(\ref{eq:up_vs_uvis_typeII}) with $c=3$.
When $\sigmaunp \rp^2/\mpl < 1$, the results given by \cite{Duffell_Haiman_MacFadyen_DOrazio_Farris2014} are also consistent with Equation~(\ref{eq:up_vs_uvis_typeII}) with $c=3$.
Although \cite{Duffell_Haiman_MacFadyen_DOrazio_Farris2014} reported that when $\sigmaunp \rp^2/\mpl >1$, the migration speed is saturated to a constant, our results and the results given by \cite{Durmann_Kley2015} show that the migration speed increase with a larger surface density of the disk.
\RED{
Equation~(\ref{eq:up_classical_typeII}) is also consistent with the results of the hydrodynamic simulations described above, when $\sigmaunp \rp^2/\mpl \ll 1$.
However, when $\sigmaunp \rp^2/\mpl > 1$, the ratio calculated by Equation~(\ref{eq:up_classical_typeII}) converges to unity, which is slower than that given by the simulations.
}

In the case of $H_0=0.03$ (lower panel of Figure~\ref{fig:up_uvis_ratio_vs_diskmass}),
\RED{only when $\sigmaunp/\rp^2/\mpl=0.02$ and $\alpha=10^{-3}$, the value of $\up/\uvis$ deviates from the line of Equation~(\ref{eq:up_vs_uvis_typeII}).
However, in most cases,} Equation~(\ref{eq:up_vs_uvis_typeII}) with $c=3$ reasonably reproduces $\up/\uvis$ given by the hydrodynamic simulations, as in the case of $H_0=0.05$.
As can be seen in Figures~\ref{fig:up_uvis_ratio_vs_diskmass}, our formula (Equation~(\ref{eq:up_vs_uvis_typeII})) can be applied, both in the cases of the massive and less massive disks.

As the previous studies \citep{Duffell_Haiman_MacFadyen_DOrazio_Farris2014,Durmann_Kley2015} have shown, the migration speed of the gap-opening planet can be faster than the gas viscous drift speed.
\cite{Durmann_Kley2015} have also shown that the migration speed of the Jupiter-sized planet is faster than the viscous velocity when $\sigmaunp \rp^2$ is larger than $0.2$ times the planet mass.
Using Equation~(\ref{eq:up_vs_uvis_typeII}), we can express the condition for $\up/\uvis >1$ as
\begin{align}
\frac{\sigmaunp \rp^2}{\mpl} > 0.2 \bracketfunc{c}{3}{-1} \bracketfunc{\hp/\rp}{0.05}{-1}.
\label{eq:crit_diskmass}
\end{align}
We should note that since the computational data is scattered in the Figure~\ref{fig:up_uvis_ratio_vs_diskmass}, we cannot accurately determine the value of $c$.
Because of it, the condition of Equation~(\ref{eq:crit_diskmass}) is not very accurate.
Nevertheless, as shown by Figure~\ref{fig:up_uvis_ratio_vs_diskmass}, we can find that $\up/\uvis > 1$ when $\sigmaunp \rp^2/\mpl > 0.1$ -- $0.2$, which is consistent with the result of \cite{Durmann_Kley2015}.

\section{An empirical model for planetary migration speed} \label{sec:empirical_model}
\subsection{Formula for the steady migration speed}
In the previous section, the torque exerted on the gap-opening planet in steady state can be described by Equation~(\ref{eq:migrate_typeII}) with $c=3$.
Here extending Equation~(\ref{eq:migrate_typeII}), we construct an empirical formula of the migration speed both for the small planet migrating in the type~I regime and the gap-opening planet.
First, let us notice that for the gap-opening planet, we may express the torque as $\Gamma = \Gamma_L/(1+0.04)$, using the fact that the Lindblad torque expected by the linear theory (Equation~(\ref{eq:GammaL_Paardekooper11})) is equal to $-3.1\Gamma_0(\rp)$ (for $f=0$) and $-2.4\Gamma_0 (\rp)$ (for $f=1/4$).
On the other hand, for the small planet with a small $K$, the torque should be obtained from the sum of $\Gamma_L$ and $\Gamma_C$.
This implies that as the gap becomes deeper, the corotation torque becomes less effective.
Hence, we can obtain the formula of the torque exerted on the planet in the form
\begin{align}
\Gamma &= \frac{\Gamma_L + \Gamma_C \exp\left({-K/K_{\rm t}}\right)}{1+0.04K},
\label{eq:formula_gamma}
\end{align}
where $\Gamma_L$ and $\Gamma_C$ are obtained from Equations~(\ref{eq:GammaL_Paardekooper11}) and (\ref{eq:GammaC_Paardekooper11}).
The exponential coefficient of $\Gamma_C$ indicates the cutoff of the corotation torque, and $K_{\rm t}$ is related to the gap depth for which the corotation torque is ineffective.
As shown in Figure~\ref{fig:torque_vs_kgap}, if $K\gtrsim 20$, the gap affects the torque.
Hence, we set $K_{\rm t}=20$.
For $K \ll K_{\rm t}$ and $0.04 K\ll 1$, the torque obtained from Equation~(\ref{eq:formula_gamma}) corresponds to that expected by the linear theory as $\Gamma_L+\Gamma_C$.
For $K \gg K_{\rm t}$ and $0.04K\gg 1$, the torque is given by $\Gamma_L/(0.04K)$.

The migration timescale of the planet is defined by
\begin{align}
\tau_{a} &= - \frac{\rp}{u_{\rm p}} \nonumber\\
&=-\frac{\rp^2\omegakp \mpl}{2\Gamma}.
\label{eq:tmig}
\end{align}
Using Equation~(\ref{eq:formula_gamma}), we obtain the migration timescale as
\begin{align}
\tmig &= -\frac{1+0.04K}{\gamma_L+\gamma_C \exp\left({-K/K_{\rm t}}\right)} \tau_0(\rp),
\label{eq:formula_ta}
\end{align}
where $\gamma_L$ and $\gamma_C$ denote $\Gamma_L/\Gamma_0(\rp)$ and $\Gamma_C/\Gamma_0(\rp)$, respectively, and
\begin{align}
\tau_{0}(R) &= \frac{R^2 \omegak \mpl}{2\Gamma_0(R)}.
\label{eq:tau0}
\end{align}

\begin{figure}
	\begin{center}
		\resizebox{0.49\textwidth}{!}{\includegraphics{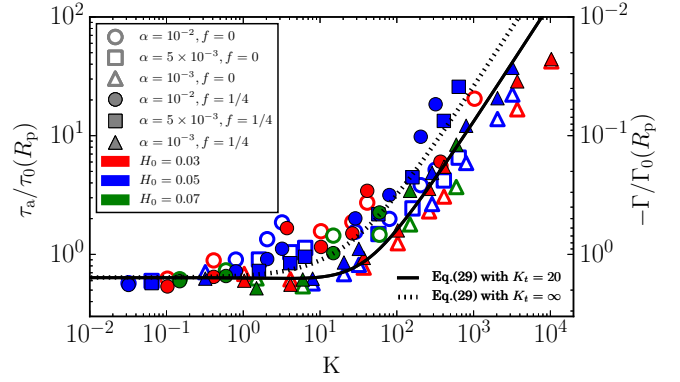}}
		\caption{
		Migration timescales versus $K$.
		The solid and dashed lines are the migration timescales given by Equation~(\ref{eq:formula_ta}) with $f=0$ in the cases of $K_{\rm t}=20$ and $K_{\rm t}=\infty$ (no corotation cutoff), respectively.
		Marks indicate values of $\alpha$ (open marks for $f=0$, and filled marks for $f=1/4$) while colors indicate values of $H_0$.
		\label{fig:tmig_vs_kgap}
		}
	\end{center}
\end{figure}
In Figure~\ref{fig:tmig_vs_kgap}, we compare the migration timescale obtained from the simulations and Equation~(\ref{eq:formula_ta}).
With $K_{\rm t}=20$, Equation~(\ref{eq:formula_ta}) can reasonably well reproduce the migration timescales given by the simulations within the factor of 2 -- 3.
In the same figure, we also plot Equation~(\ref{eq:formula_ta}) with $K_{\rm t}=\infty$, which does not consider the corotation cutoff.
In the cases with the relatively high viscosity (i.e., $\alpha=10^{-2}$), Equation~(\ref{eq:formula_ta}) with $K_{\rm t}=\infty$ gives a better fit for the migration timescale in the range of $10<K<10^{3}$, which implies that the corotation torque still affects the planetary migration even if the planet forms the gap.
The effect of the corotation torque on the planetary migration is discussed in Section~\ref{sec:corotations}.

\subsection{Time variation of the migration speed} \label{eq:timevariation_tmig}
So far, we discuss the migration speed in steady state.
In this subsection, we discuss the time variation of the migration speed, before the steady state is reached.
According to \cite{Kanagawa2017b}, the timescale that the gap becomes a steady state structure ($t_{\rm gap}$) can be scaled by $(\mpl/\mstar)(\hp/\rp)^{-7/2}\alpha^{-3/2}t_0$ (see also Appendix~\ref{sec:gap_structures}).
The timescale after which the torque is saturated can be assumed to be similar to $t_{\rm gap}$.
When $t>t_{\rm gap}$, the migration timescale corresponds to the migration timescale in steady state which is given by Equation~(\ref{eq:formula_ta}).
Assuming a simple exponential time variation with $t_{\rm gap}$, we may describe time variations of the migration timescale of the planet, as
\begin{align}
\tau_{a} (t) &= \frac{1+0.04K \left(1-e^{-t/t_{\rm gap}} \right)}{\gamma_L + \gamma_C \exp\left(-K/K_{\rm t} \right)} \tau_0(\rp),
\label{eq:formula_ta_with_t}
\end{align}
where $t_{\rm gap}$ may be given by
\begin{align}
t_{\rm gap} &= 2.4\times 10^{3} \bracketfunc{\mpl/\mstar}{10^{-3}}{} \bracketfunc{\hp/\rp}{0.05}{-7/2} \bracketfunc{\alpha}{10^{-3}}{-3/2} t_0.
\label{eq:gap_opening_time}
\end{align}

\begin{figure}
	\begin{center}
		\resizebox{0.49\textwidth}{!}{\includegraphics{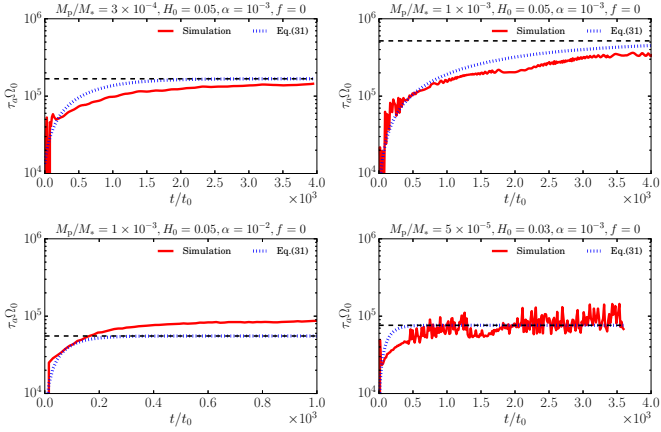}}
		\caption{
		Time variations of the migration timescale obtained from the hydrodynamic simulations (red solid) and Equation~(\ref{eq:formula_ta_with_t})(blue dotted).
		The horizontal dashed lines indicates the migration timescale in steady state given by Equation~(\ref{eq:formula_ta}).
		\label{fig:evo_tmig_wmodel}
		}
	\end{center}
\end{figure}
We show comparisons between the migration timescales given by the simulations and Equation~(\ref{eq:formula_ta_with_t}) in several cases, in Figure~\ref{fig:evo_tmig_wmodel}.
In most cases, Equation~(\ref{eq:formula_ta_with_t}) gives a reasonable fit of the time variation of $\tau_{a}$.

\section{Effects of the corotation torque} \label{sec:corotations}
The corotation torque does not need to be negligible during the migration of the gap-opening planet, even if the gap is relatively deep, namely such that $K\gtrsim 20$.
If $\mpl/\mstar \gtrsim (\hp/\rp)^3$, the nonlinear effects of the corotation torque are important.
In this section, we discuss the effects of the corotation torque on the planetary migration.
When the planet forms a relatively shallow gap, the very fast type~III migration can occur \citep{Masset_Papaloizou2003}.
Moreover, as shown in \cite{Masset_DAngelo_Kley2006}, the corotatoin torque is significantly boosted by the nonlinear effects, which can reduce the migration speed \citep{Duffell2015b}.

\begin{figure}
	\begin{center}
		\resizebox{0.32\textwidth}{!}{\includegraphics{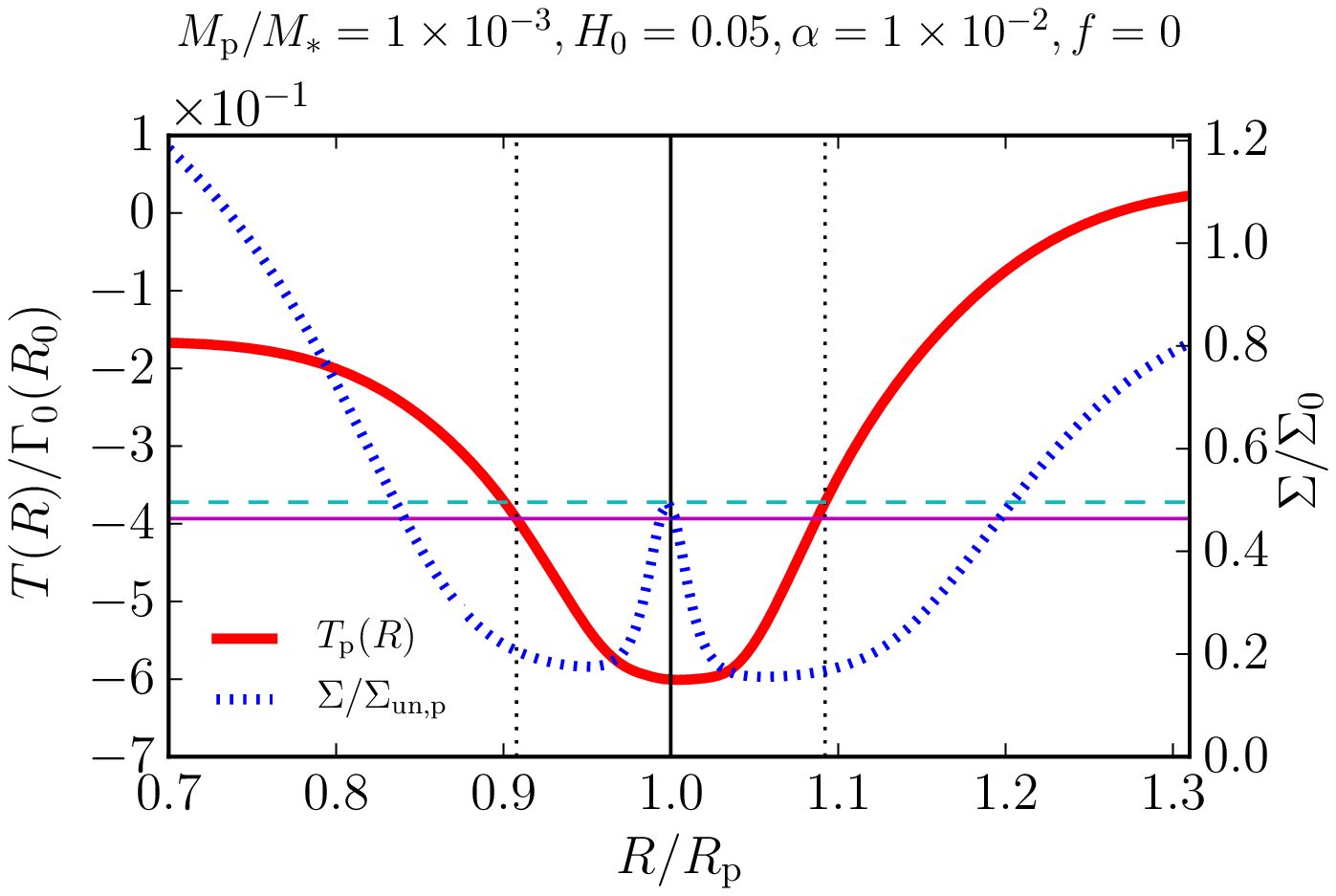}}
		\resizebox{0.32\textwidth}{!}{\includegraphics{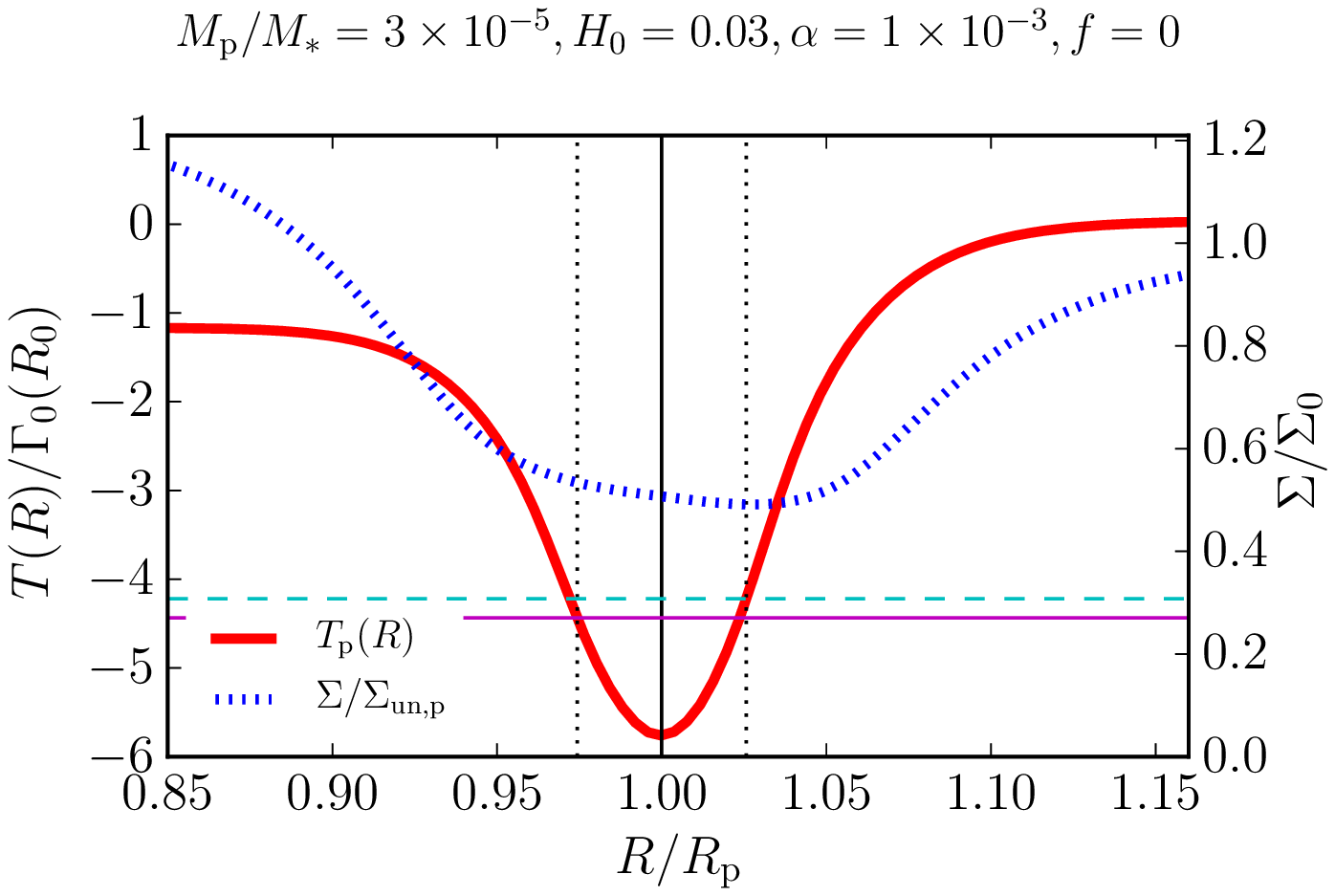}}
		\resizebox{0.32\textwidth}{!}{\includegraphics{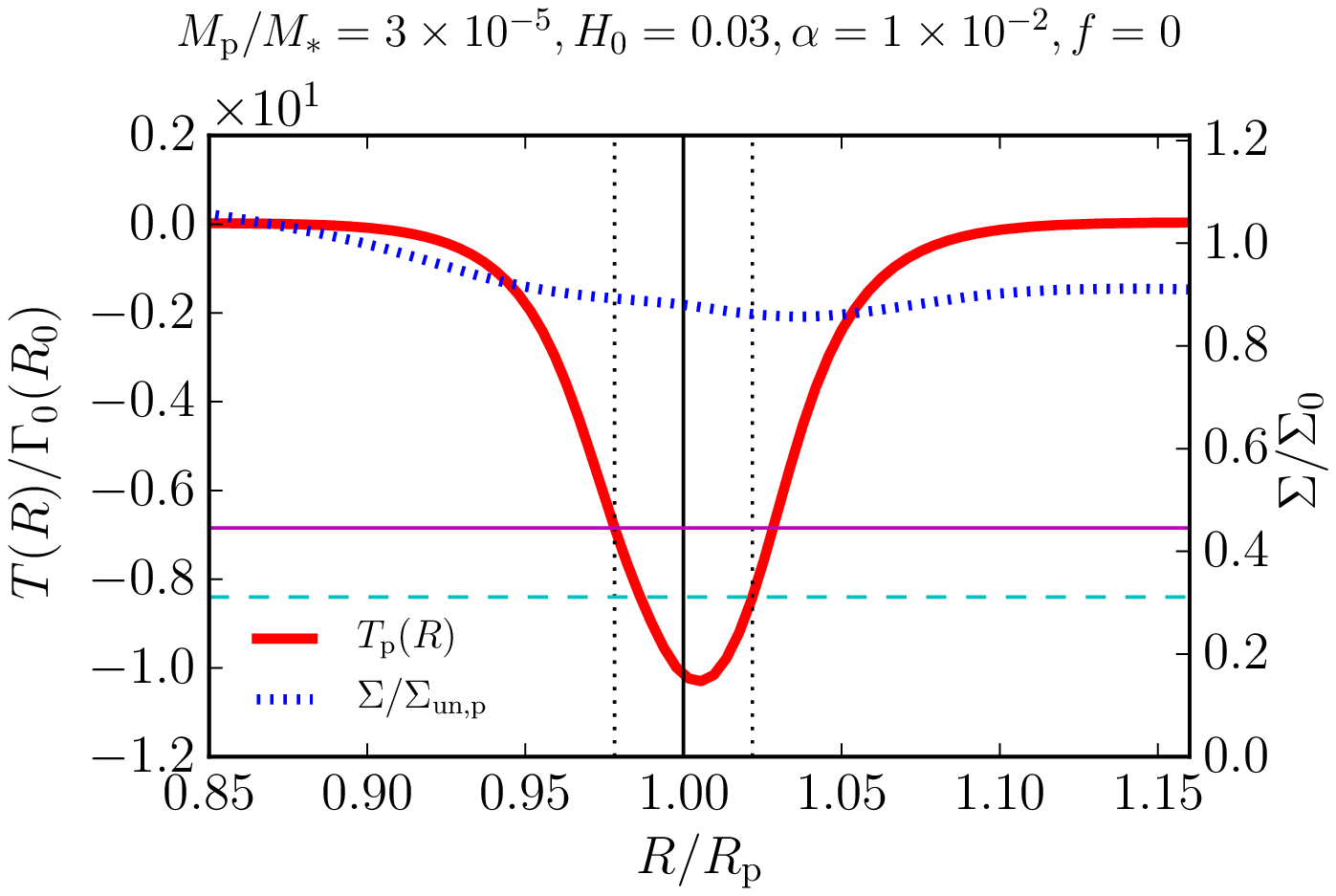}}
		\caption{
		The distributions of cumulative torques exerted on the planet by the disk (solid line), and the azimuthally averaged surface density (dashed line).
		The planet mass, the disk aspect ratio, and the viscosity in the upper panel are $\mpl/\mstar=10^{-3}$, $H_0=0.05$, and $\alpha=10^{-2}$, respectively.
		In the middle and lower panels, the planet mass and disk aspect ratio are $\mpl/\mstar=3\times 10^{-5}$ and $H_0=0.03$, respectively, and the values of $\alpha$ is $10^{-3}$ in the middle panel and $\alpha=10^{-2}$ in the lower panel.
		The vertical thin solid line denotes the planet's location ($\rp$) and the two vertical thin dashed lines denote the locations of $\rp-\rhill$ and $\rp+\rhill$ from the left, respectively.
		The solid and dashed thin horizontal lines denote $\tp(\rp-\rhill)$ and $\tp(\rp+\rhill)$, respectively.
		\label{fig:cumulative_torques}
		}
	\end{center}
\end{figure}
Figure~\ref{fig:cumulative_torques} shows distributions of the cumulative torque (Equation~(\ref{eq:cumulative_torque})) for three particular sets of parameters.
For reference, we also plot the azimuthally averaged surface density.
The corotation torque is related to the horseshoe drag and exerted from the inside of the horseshoe region.
In the case where the nonlinear effects are significant (and then $\mpl/\mstar > (\hp/\rp)^3$), the width of the horseshoe region is approximately equal to the Hill radius, $\rhill=(\mpl/(3\mstar))^{1/3}$ \citep{Masset_DAngelo_Kley2006,Paardekooper_Papaloizou2009}.
Although not examining the stream lines, we can roughly estimate the effects of the corotation torque by measuring the torque from the inside of the horseshoe region.
\RED{
The torque coming from the inside of the horseshoe region ($\rp-\rhill<R<\rp+\rhill$) is given by $\int^{\rp+\rhill}_{\rp-\rhill} (d\Gamma/dR)dR' = \tp(\rp-\rhill)-\tp(\rp+\rhill)$.
}
When the large planet forms the relatively deep gap as in the upper panel of Figure~\ref{fig:cumulative_torques}, within the horseshoe region, the cumulative torques coming from the inner and outer parts of the disk almost cancel out each other.
In this case, the corotation torque is negligible as compared with the Lindbald torque.

As \cite{Masset_Papaloizou2003} have shown, the fast inward migration (type~III migration) can occur when the gap is relatively shallow, because of the negative corotation torque.
As can be seen in the middle panel of Figure~\ref{fig:cumulative_torques}, the relatively shallow gap is formed.
In this case, the value of $\tp(\rp-\rhill) - \tp (\rp+\rhill)$, which can be considered as the torque exerted on the planet by the gas within the horseshoe region, is negative.
This negative corotation torque promotes the inward migration, which could be treated as the type~III migration.
In the lower panel of Figure~\ref{fig:cumulative_torques}, on the other hand, the planet mass and the disk aspect ratio are the same as in the middle panel, but the value of $\alpha$ ($=10^{-2}$) is larger than that in the middle panel.
Since the viscosity is larger, the gap in the lower panel is much shallower than that in the middle panel.
\RED{
In this case, since $\tp(\rp-\rhill) - \tp(\rp+\rhill) > 0$, we can find that the significant positive torque is generated in the horseshoe region.
}
Because of this enhancement of the corotation torque, the torque exerted on the planet is positive.
This boost of the positive torque could be explained by the nonlinear effects of the corotation torque discussed by \cite{Masset_DAngelo_Kley2006,Duffell2015b}.

\begin{figure}
	\begin{center}
		\resizebox{0.48\textwidth}{!}{\includegraphics{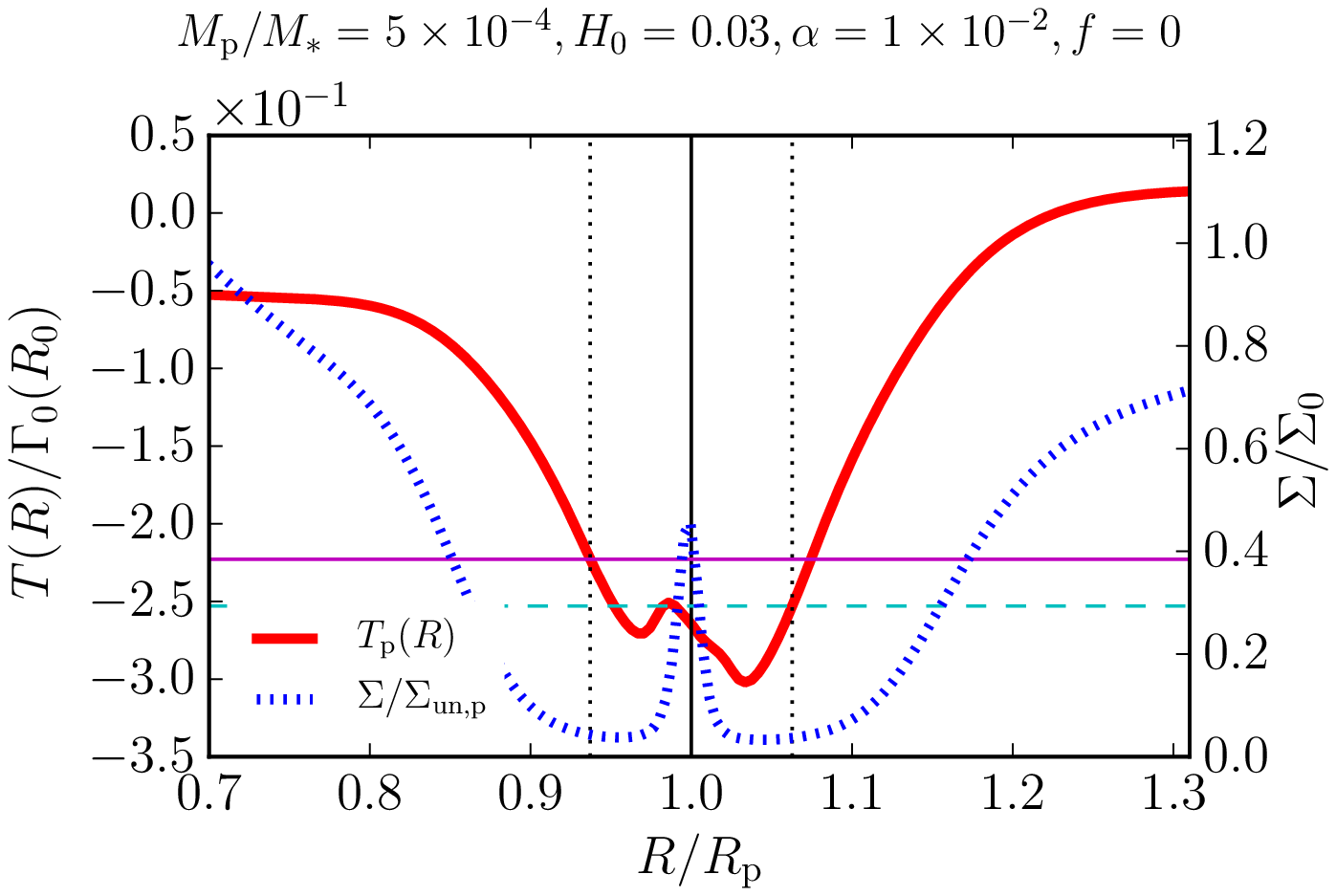}}
		\resizebox{0.48\textwidth}{!}{\includegraphics{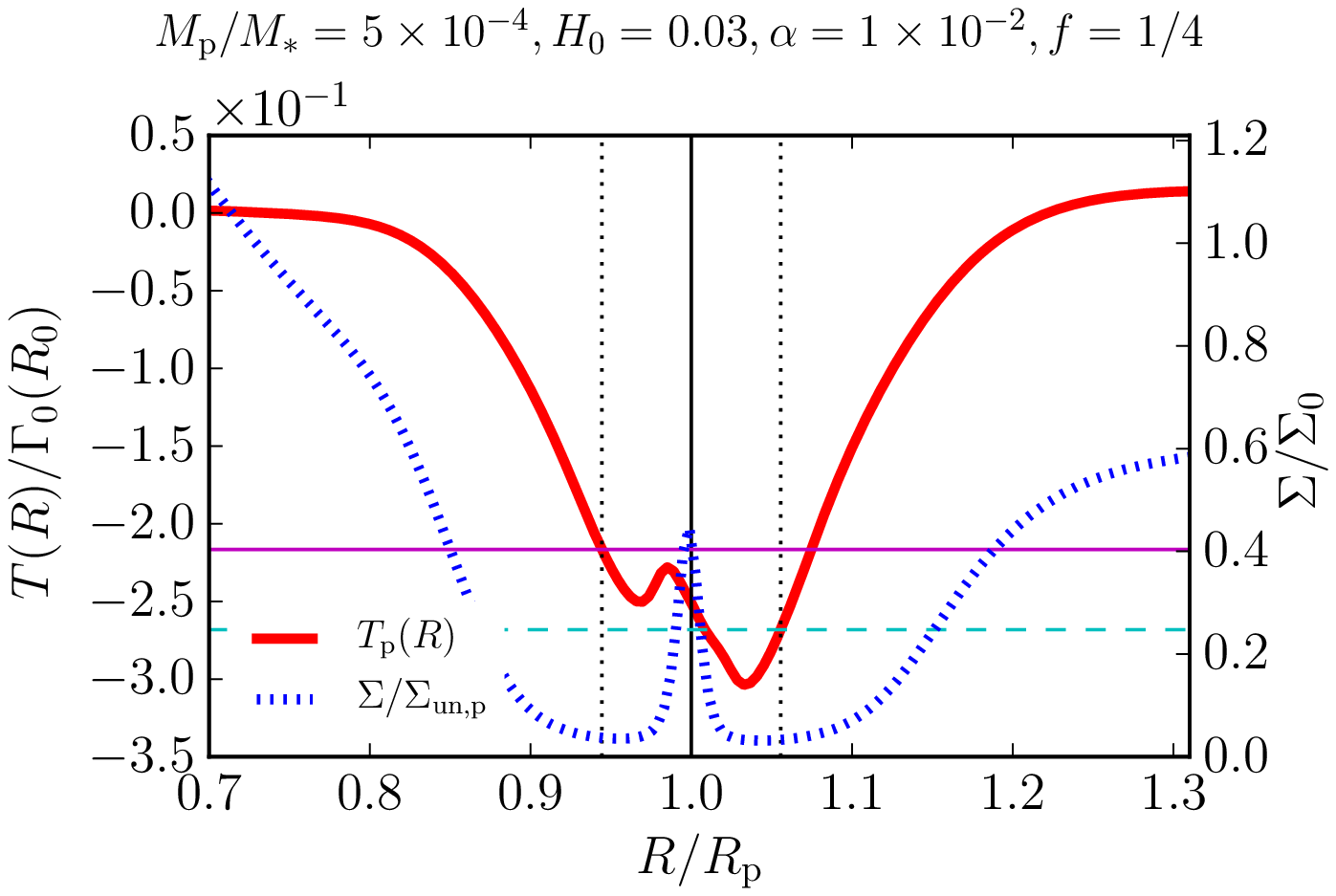}}
		\caption{
		The same as Figure~\ref{fig:cumulative_torques}, but in the cases of $\mpl/\mstar=5\times 10^{-4}$, $H_0=0.03$, and $\alpha=10^{-2}$.
		The flaring indexes are set to be $0$ (upper) and $1/4$ (bottom), respectively.
		\label{fig:cumulative_torques2}
		}
	\end{center}
\end{figure}
\RED{
Now we show another example of the boost of the positive corotation torque in the case when the gap is relatively deep.
Figure~\ref{fig:cumulative_torques2} shows the distributions of the cumulative torques with $f=0$ and $f=1/4$, when $\mpl/\mstar=5\times 10^{-4}$, $H_0=0.03$ and $\alpha=10^{-2}$.
In these cases, the gaps are relatively deep (i.e., $\sigmamin/\sigmaunp \simeq 0.1$).
Nevertheless, as can be seen in the figure, the distributions of the cumulative torques are asymmetric in the inner and outer parts of the horseshoe region, which generate the strong positive torque.
This positive torque slows down the planetary migration.
Moreover, this positive torque depends on the flaring index.
As can be seen in Figure~\ref{fig:cumulative_torques2}, when $f=1/4$, the positive torque originated from the inside of the horseshoe region is so strong that the planet migration is almost halted.
For a more accurate model of the planetary migration, comprehensive understanding of the nonlinear corotation torque is required.
}

Although we adopt the locally isothermal equation of state in this paper, in the case of a non-isothermal equation of state, the corotation torque becomes much effective than that in the locally isothermal case, as shown by \cite{Paardekooper_Baruteau_Crida_Kley2010}.
In this case, the migration of the planet with a shallow gap may be significantly slow-down by the non-isothermal effects, in the same way as the small planet in the linear regime.
Even in such case, the torque may be roughly proportional to the surface density at the bottom of the gap, as in Figure~\ref{fig:torque_vs_smin}.
If so, Equation~(\ref{eq:formula_ta}) may be valid even in the non-isothermal case, with use of $\Gamma_L$ and $\Gamma_C$ in the non-isothermal case.

\section{Summary} \label{sec:summary}
We have investigated the migration of the planet in the protoplanetary disk by performing over hundred runs of two-dimensional hydrodynamic simulations.
Our results are summarized as follows:
\begin{enumerate}
  \item We found that the torque exerted on the gap-opening planet is proportional (with some scatter) to the surface density at the bottom of the gap (Figure~\ref{fig:torque_vs_smin}), \RED{when the mass of the planet is smaller than Jupiter and the disk parameters have their standard values: the disk aspect ratio and viscosity are in the ranges of $0.03 < H_0 < 0.07$ and $10^{-2} < \alpha < 10^{-3}$, respectively. Hence, the migration of the gap-opening planet slows down simply because of the decrease in the gas surface density at the bottom of the gap.}
  \item As shown in Figure~\ref{fig:torque_vs_kgap}, the transition between the type~I migration and the migration of the gap-opening planet occurs around $K=20$, which can be referred to as the gap formation criterion in terms of the planetary migration.
  \item The migration speed of the gap-opening planet can be faster than the gas viscous accretion speed if the disk is massive enough namely $\sigmaunp \rp^2/\mpl > 0.1$ -- $0.2$ (Figure~\ref{fig:up_uvis_ratio_vs_diskmass}). This result is consistent with \cite{Durmann_Kley2015}.
  \item Considering the reduction of the torque due to the gap formation, we present the simple model of the planetary migration (Equation~(\ref{eq:formula_ta})), which can reasonably well reproduce the migration speed given by the hydrodynamic simulations (Figure~\ref{fig:tmig_vs_kgap}). \RED{This formula can be applied both for the massive disk (disk-dominated case) and less massive disk (planet-dominated case), and it can be useful to be incorporated into the population synthesis calculation.}
\end{enumerate}

\acknowledgements
We would like to thank Professor Wilhelm Kley at University of T\"{u}bingen for fruitful discussion.
This work was supported by the Polish National Science Centre MAESTRO grant DEC- 2012/06/A/ST9/00276.
KDK was also supported by JSPS Core-to-Core Program ``International Network of Planetary Sciences''.
Numerical computations were carried out on the Cray XC30 at the Center for Computational Astrophysics, National Astronomical Observatory of Japan.

\appendix
\section{Gap structures induced by migrating planets} \label{sec:gap_structures}
The gap depth and width have been investigated by the hydrodynamic simulations, for various planet masses, aspect ratios, and viscosities \citep{Duffell_MacFadyen2013,Fung_Shi_Chiang2014,Kanagawa2015b,Kanagawa2016a,Kanagawa2017b}.
However, these results are given by assuming a fixed orbit of the planet.
Here we confirm the validity of these formula in the case in which the planet migrates.

The gap depth in steady state is obtained from Equation~(\ref{eq:smin}).
\begin{figure}
	\begin{center}
		\resizebox{0.49\textwidth}{!}{\includegraphics{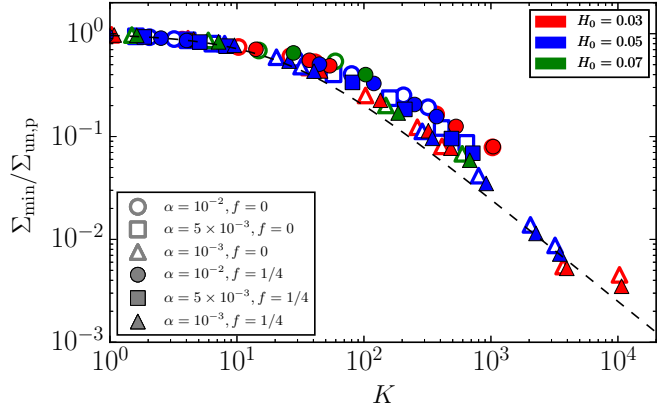}}
		\caption{
		The surfaces density at the bottom of the gap given by the hydrodynamic simulations.
		The dashed line indicates the density given by Equation~(\ref{eq:smin}).
		Marks indicate values of $\alpha$ (open marks for $f=0$, and filled marks for $f=1/4$) while colors indicate values of $H_0$.
		\label{fig:smin_vs_kgap}
		}
	\end{center}
\end{figure}
In Figure~\ref{fig:smin_vs_kgap}, we show the surface density at the bottom of the gap given by the simulations with the migrating planet, at $3000t_0$\footnote{
In few cases, the values at $\rp=0.7R_0$ are treated as the stationary values, as mentioned in Section~\ref{subsec:timevar_torque}.
}.
As can be seen in the figure, Equation~(\ref{eq:smin}) reasonably well reproduce $\sigmamin$ given by the simulations.
Note that when $\alpha=10^{-2}$, in the range of $10^{2}\gtrsim K \gtrsim 10^{3}$, the value of $\sigmamin$ given by the simulations is slightly larger than that given by Equation~(\ref{eq:smin}).
However, even in these cases, the discrepancy between Equation~(\ref{eq:smin}) and that given by the simulations is a factor of two.

The formula of the gap width is provided by \citep{Kanagawa2016a,Kanagawa2017b}.
They measure a gap width as a radial width of a region where the surface density is smaller than a density threshold, $\rhosurf_{\rm th}$.
In this case, the gap width in steady state is given by
\begin{align}
\frac{\wgap}{\rp} &= \left(0.5 \frac{\rhosurf_{\rm th}}{\sigmaunp}+0.16\right)K'^{1/4},
\label{eq:width_kprime}
\end{align}
where
\begin{align}
K'&=\bracketfunc{\mpl}{\mstar}{2}\bracketfunc{\hp}{\rp}{-3}\alpha^{-1}.
\label{eq:kprime}
\end{align}
\begin{figure}
	\begin{center}
		\resizebox{0.49\textwidth}{!}{\includegraphics{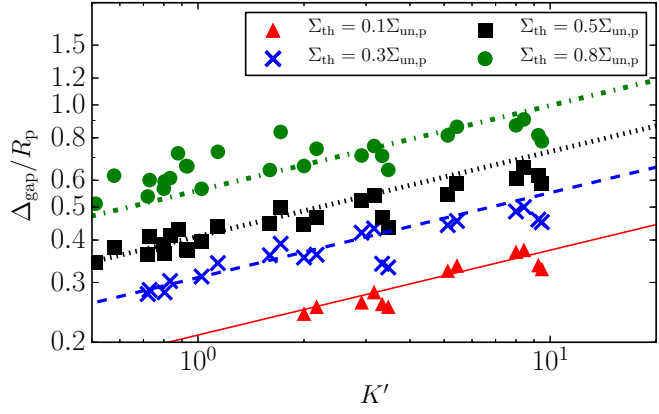}}
		\caption{
		Gap widths given by the hydrodynamic simulations.
		The lines denote the widths obtained from Equation~(\ref{eq:width_kprime}) for $\Sigma_{\rm th}/\sigmaunp = 0.1$, $0.3$, $0.5$, and $0.8$ from the bottom, respectively.
		The triangles, crosses, squares, and circles indicate the width measured from the simulations, for $\Sigma_{\rm th}/\sigmaunp = 0.1$, $0.3$, $0.5$, and $0.8$, respectively.
		\label{fig:width_vs_kprime}
		}
	\end{center}
\end{figure}
Figure~\ref{fig:width_vs_kprime} shows the gap widths given by the simulations for $\rhosurf_{\rm th}/\sigmaunp=0.1,0.3,0.5$ and $0.8$.
As can be seen in the figure, the gap widths obtained from Equation~(\ref{eq:width_kprime}) is in reasonably agreement with the widths measured in the simulations, as in the case of the gap depth.

The gap-opening timescale can be estimated by $t_{\rm gap} = (\wgap/2\rp)^2 \rp^2/\nu$.
Using $\wgap$ measured by $\rhosurf_{\rm th}/\sigmaunp = 0.5$, we obtain $t_{\rm gap} = 0.1K'^{2} \rp^2/\nu$ \citep{Kanagawa2017b}.
In our survey, the migration timescale is $\tau_{a} \gtrsim 2\times 10^{4} t_0$, which is longer than the gap-opening time in most cases.
Hence, the planet can hold the gap depth and width given by Equations~(\ref{eq:smin}) and (\ref{eq:width_kprime}) during its' migration, as can be seen in Figures~\ref{fig:smin_vs_kgap} and \ref{fig:width_vs_kprime}.



\end{document}